k
\documentclass[11pt,letter]{emulateapj}
\usepackage{graphicx}
\usepackage{amsmath}
\usepackage{amssymb}
\usepackage{hyperref}

\shorttitle{300-480\,MHz Polarization Survey}
\shortauthors{Wolleben et. al.}

\begin{document}

\title{The Global Magneto-Ionic Medium Survey: Polarimetry of the Southern Sky
from 300 to 480 MHz}

\author{M. Wolleben\altaffilmark{1,2}, T.L. Landecker\altaffilmark{1}, E.
Carretti\altaffilmark{3,4}, J.M. Dickey\altaffilmark{5}, A.
Fletcher\altaffilmark{6},  N.M. McClure-Griffiths\altaffilmark{7}, D.
McConnell\altaffilmark{3}, A.J.M. Thomson\altaffilmark{7}, A.S.
Hill\altaffilmark{8,1,9}, B.M.
Gaensler\altaffilmark{10,11}, J.-L. Han\altaffilmark{12,13,14},  M.
Haverkorn\altaffilmark{15}, J.P. Leahy\altaffilmark{16},  W.
Reich\altaffilmark{17},  A.R. Taylor\altaffilmark{18,19}}

\altaffiltext{1}{National Research Council Canada, Herzberg Research Centre for
Astronomy and Astrophysics, Dominion Radio Astrophysical Observatory, P.O. Box
248,  Penticton, British Columbia, V2A 6J9, Canada}         

\altaffiltext{2}{Skaha Remote Sensing, 3165 Juniper Drive, Naramata, British
Columbia, V0H 1N0, Canada}

\altaffiltext{3}{CSIRO Astronomy \& Space Science, P.O. Box 76, Epping, New
South Wales 1710, Australia}

\altaffiltext{4}{INAF - Istituto di Radioastronomia, via P. Gobetti 101, I-40129
Bologna, Italy}

\altaffiltext{5}{University of Tasmania, School of Mathematics and Physics, 
Hobart, Tasmania 7001, Australia}

\altaffiltext{6}{School of Mathematics, Statistics and Physics, Newcastle
University, Newcastle upon Tyne, NE1 7RU, United Kingdom}

\altaffiltext{7}{Research School of Astronomy and Astrophysics, Australian
National University, Canberra, Australian Capital Territory 2611, Australia}

\altaffiltext{8}{Department of Physics and Astronomy, University of British
Columbia, 6224 Agricultural Road, Vancouver, British Columbia, V6T 1Z1, Canada}

\altaffiltext{9}{Space Science Institute, Boulder, CO USA, 80301}

\altaffiltext{10}{Sydney Institute for Astronomy, School of Physics, University
of Sydney, Sydney, New South Wales 2006, Australia} 

\altaffiltext{11}{Dunlap Institute for Astronomy and Astrophysics, University of
Toronto, 50 St. George Street, Toronto, M5S 3H4, Canada}

\altaffiltext{12}{National Astronomical Observatories, CAS, Jia-20
DaTun Road, Chaoyang District, Beijing 100101, China}

\altaffiltext{13}{School of Astronomy and Space Sciences, University of the
Chinese Academy of Sciences, Beijing 100049, China}

\altaffiltext{14}{CAS Key Laboratory of FAST, NAOC, Chinese Academy of Sciences,
Beijing 100101, China}

\altaffiltext{15}{Department of Astrophysics/IMAPP, Radboud University Nijmegen,
P.O. Box 9010, NL-6500 GL, Nijmegen, the Netherlands}

\altaffiltext{16}{Jodrell Bank Centre for Astrophysics, Alan Turing Building,
School of Physics and  Astronomy, University of Manchester, Oxford Road,
Manchester, M13 9PL, United Kingdom}

\altaffiltext{17}{Max-Planck-Institut f\"ur Radioastronomie, Auf dem H\"ugel 69,
D-53121 Bonn, Germany}

\altaffiltext{18}{Department of Astronomy, University of Cape Town, Rondenbosch
7701, Republic of South Africa}

\altaffiltext{19}{Department of Physics, University of the Western Cape,
Republic of South Africa}

\begin{abstract}     
{ Much data on the Galactic polarized radio emission has been gathered in the
last five decades. All-sky surveys have been made, but only in narrow, widely
spaced frequency bands, and the data are inadequate for the characterization of
Faraday rotation, the main determinant of the appearance of the polarized radio
sky at decimetre wavelengths. We describe a survey of polarized radio emission
from the Southern sky, aiming to characterize the magneto-ionic medium,
particularly the strength and configuration of the magnetic field. This work is part of the Global Magneto-Ionic Medium Survey (GMIMS). We have
designed and built a feed and receiver covering the band 300 to 900\,MHz for the
CSIRO Parkes 64-m Telescope.  We have surveyed the entire sky between
declinations $-90^{\circ}$ and $+20^{\circ}$. We present data covering 300 to
480\,MHz with angular resolution $81'$ to $45'$. The survey intensity scale is
absolutely calibrated, based on measurements of resistors at known temperatures
and on an assumed flux density and spectral index for Taurus A. Data are
presented as brightness temperatures. We have applied Rotation Measure Synthesis
to the data to obtain a Faraday depth cube of resolution
5.9\,${\rm{rad}}\thinspace{\rm{m}}^{-2}$, sensitivity of 60\,mK of polarized
intensity, and angular resolution $1.35^{\circ}$. The data presented in this paper are available at the Canadian Astronomy Data Centre.}  
\end{abstract}


\section{Introduction}
\label{intro}

The first detections of linearly polarized components of the Galactic radio
emission were made over fifty years ago \citep{west62,wiel62} at frequencies
near 400\,MHz. These discoveries firmly established the synchrotron mechanism as
the source of the Galactic non-thermal emission and confirmed the existence of
magnetic fields in the Milky Way. The earliest papers commented on the role of
Faraday rotation in shaping the appearance of the polarized sky, which was
distinctly different from the sky in total intensity.  Surveys at other
frequencies soon followed ({\it{e.g.}}
\citealp{berk63,math65,math66,bing66,wilk73,bake74}). The most comprehensive of
these are surveys made during the 1960s with the Dwingeloo 25-m Telescope,
reduced in a systematic way and published by \citet{brou76}. These surveys, at
408, 465, 610, 820, and 1411\,MHz, cover the entire Northern sky, albeit with
sparse spatial sampling. Absolutely calibrated and carefully processed, they are
considered among the best representations of the polarized sky available.

Faraday rotation of the polarized signal from extragalactic sources was detected
by \citet{cp62}, and systematic observations over large areas were soon used to
generate models of the large-scale Galactic magnetic field
(\citealp{gard63,seym66,sima79,brow10,han17}). Observations at widely spaced
frequencies had proved adequate to determine the Rotation Measure (RM) of
extragalactic sources, and it was tacitly assumed that the same would be true of
the Galactic synchrotron emission. On this basis, \citet{spoe84} calculated the
RM of the Galactic emission based on the data at 408, 465, 610, 820 and
1411\,MHz of \citet{brou76}.

In contrast, the complexity of Faraday effects in a medium where emission and
Faraday rotation are mixed had already been demonstrated by \citet{burn66}. Burn
showed that, when synchrotron emission and Faraday rotation are present in the
same volume, as must often be the case in the Galaxy, ``Rotation Measure'' is
not a meaningful concept and measurements in many closely spaced frequency
channels are required to fully characterize Faraday rotation. However, the
technology of the 1960s was inadequate to collect or to process signals with the
necessary bandwidth and frequency resolution. 

All-sky surveys at single frequencies ({\it{e.g.}}
\citealp{brou76,woll06,test08}) and aperture-synthesis surveys in the Galactic
plane (\citealp{have06a,land10}) have provided
two-dimensional mapping of the Galactic magnetic field.  In combination with
other tracers, such as Faraday rotation towards point sources
\citep{HanManchester:2006,BrownHaverkorn:2007,TaylorStil:2009}, they have
contributed to reconstructions of the three-dimensional structure of the
Galactic magnetic field \citep{SunReich:2008,JanssonFarrar:2012}.

Today, advances in antennas, receivers, and digital signal processing have made
polarimetry possible over wide bands with many frequency channels. The technique
of Rotation Measure Synthesis \citep{bren05}, drawing on the concepts of
\citet{burn66}, has been developed and applied to aperture synthesis
observations \citep{debr05}. Our Global Magneto-Ionic Medium Survey (GMIMS)
exploits these opportunities. With GMIMS we set out to map the polarized radio
emission from the entire sky, in the Northern and Southern hemispheres, using
large single-antenna radio telescopes, covering the entire frequency range 300
to 1800\,MHz \citep{woll09}, and to use RM Synthesis to analyze the data. The
frequency band has been notionally divided in three, 300 to 800\,MHz, 800 to
1300\,MHz, and 1300 to 1800\,MHz, to define the Low, Mid, and High GMIMS bands.
Since any one telescope can see just over half the sky, we obviously required
North and South surveys in each of the three bands. We developed the techniques
for GMIMS with a survey of the High GMIMS band made with the John A. Galt
Telescope at the Dominion Radio Astrophysical Observatory (diameter 26\,m) and
we have now completed a spectropolarimetric survey of the Northern sky covering
1280 to 1750\,MHz using that telescope. Scientific results from this survey have
been published by \citet{woll10b}, \citet{sun15} and \citet{hill17}, and the
survey itself will be described in a forthcoming paper (M. Wolleben et al, in
preparation 2019). We have also used the CSIRO Parkes 64-m Telescope to make two
surveys of the Southern sky, one in the High band and one in the Low. In this
paper we describe the Parkes survey  in the Low Band.

The motivation for the GMIMS project is our conviction that the magnetic field
is an important energy-carrying constituent of the interstellar medium (ISM).
\citet{Ferriere:2001vr} envisages the ISM with three principal constituents: gas
(in cold [$\sim 10^2\,K$], warm [$\sim 10^4\,K$], and hot [$\sim 10^6\,K$]
thermally-stable ``phases''), magnetic fields, and cosmic rays. On large scales
the energy densities of these contituents are in approximate equipartition.
Nevertheless, there are substantial local and scale-dependent deviations from
equilibrium \citep{WolfireMcKee:2003,JoungMac-Low:2009,HeilesHaverkorn:2012} and
we cannot understand any one constituent of the ISM in isolation. For example,
the presence of the magnetic field and other nonthermal pressure components can
fundamentally change the character of the ISM from one dominated by hot gas with
embedded warm and cold clouds \citep{McKeeOstriker:1977,LiOstriker:2015} to one
dominated by warm gas with embedded cold clouds and hot supernova remnants
\citep{SlavinCox:1993,AvillezBreitschwerdt:2004,AvillezBreitschwerdt:2005,
GresselElstner:2008}. Overall, thermal and non-thermal pressures provide the
vertical support which keeps the ISM in hydrostatic equilibrium
\citep{BoularesCox:1990,PiontekOstriker:2007,GresselElstner:2008,
OstrikerMcKee:2010,HillJoung:2012} and the multiphase turbulent cascade in the
ISM on all scales is partly controlled by magnetism
\citep{ArmstrongRickett:1995,MinterSpangler:1996,ChepurnovLazarian:2010}.

Observational studies of the three-dimensional distribution of the ISM gas are
in a far more advanced state than those of the magnetic field and cosmic rays.
There are now all-sky, kinematically-resolved surveys of \ion{H}{1} emission
with sub-degree angular resolution
\citep{KalberlaBurton:2005,McClure-GriffithsPisano:2009,KerpWinkel:2011,
Ben-Bekhti:2016} as well as arcminute-resolution surveys of smaller areas,
completed
\citep{TaylorGibson:2003,McClure-GriffithsDickey:2005,PeekHeiles:2011,mccl12,
PeekBabler:2018} and planned \citep{DickeyMcClure-Griffiths:2013}. There is also
an all-sky, kinematically-resolved survey of H$\alpha$ emission from the warm
ionized ISM \citep{HaffnerReynolds:2003,HaffnerReynolds:2010a}{\footnote{
http://www.astro.wisc.edu/wham-site/wham-sky-survey}}. Through the velocity
dimension these surveys provide information on the distribution, structure, and
kinematics of neutral and ionized gas in the Milky Way. With GMIMS, we aim to
provide an all-sky counterpart to these surveys by mapping polarized emission
from the magneto-ionized ISM, with information on the third dimension provided
by Faraday depth. While Faraday depth is not a direct proxy for distance, it can
provide information on the three-dimensional structure of the magnetic field,
information that is not accessible in any other way.

Dust intermingles with the ISM gas. It is small in mass fraction, amounting to
${\sim}1\%$ of the ISM, but crucial for interstellar chemistry and star
formation. Dust grains are aligned by magnetic fields and dust observations play
a major role in studies of the Galactic magnetic field ({\it{e.g.}}
\citealp{plan16}). Dust polarization traces the magnetic field in the plane of
the sky while Faraday effects trace the field in the line of sight: the two
kinds of observations are complementary.

In this paper we describe a polarization survey of the Southern sky using the
Parkes 64-m Telescope. The initial goal was to map the sky from 300 to 900 MHz,
but strong radio-frequency interference (RFI) prevented this. We present data
between 300 and 480\,MHz covering 2.68$\pi$ steradians, 67.1\% of the whole sky.
Section~\ref{focus} describes the feed and receiver designed and built for this
survey. Survey observations are described in Section~\ref{obs}. In
Section~\ref{calib} we describe the methods used to calibrate the survey in
terms of absolute standards of noise. Data processing is the subject of
Section~\ref{process}. Section~\ref{data_qual} probes the quality of the survey
data through comparisons with existing data. Results are presented in
Section~\ref{results} and are discussed in Section~\ref{disc}.

\section{Feed and Receiver}
\label{focus}

The target band, 300 to 900 MHz, was chosen as a balance between a need to reach
low frequencies to achieve good resolution in Faraday depth, limitations imposed
by RFI at the Parkes observatory, the 1-GHz input bandwidth of the available
digital signal processor, and the feasibility of building an appropriate feed to
collect the signals.  Neither feed nor receiver was available for the telescope
covering our chosen band; this section of the paper describes their design and
implementation.

\subsection{Feed Design}

The specifications for the feed called for constant illumination of the 64-m
reflector over the band 300 to 900\,MHz, and the delivery of left-hand and
right-hand circular polarization (which we denote by $L$ and $R$) to the
receiver and the digital signal processor.  The Parkes Telescope has a diameter,
$D$, of 64 m and a focal length, $f$, of 26 m. $f/D = 0.41$ and the opening
angle of the reflector as seen from the focus is $126^{\circ}$. 

\begin{figure}
\centering
\includegraphics[width=1.0\columnwidth,bb= 0 70 540 560,clip=]{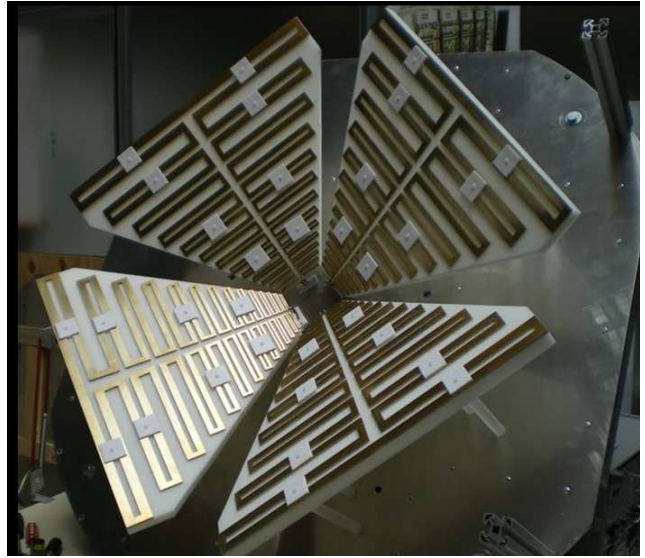}
 \caption{The feed, designed and built for this survey. Four petals, each a
 log-periodic array of folded dipoles, are supported above a ground plane of
 dimensions $1.2{\times}1.2\,{\rm{m}}$.}
 \label{feed}
\end{figure}

Our design is based on the Eleven Feed, an invention of \citet{kild05}, which
has been shown to be able to meet a variety of wideband needs
\citep{olss06,yang11}. The basic element of the Eleven Feed is a pair of
parallel half-wave dipoles above a ground plane: such a feed can easily be built
to provide a circularly symmetric illumination of a reflector of opening angle
${\sim}120^{\circ}$. In the Eleven Feed each dipole is expanded to become an
approximately log-periodic array of folded dipoles fed by a twin-wire
transmission line. We refer to this structure as a {\it{petal}}. Our feed, shown
in Figure~\ref{feed}, consists of four petals, each connected to the receiver at
the short-dipole end. For operating wavelength, $\lambda$, the dipoles of length
${\sim}0.5{\lambda}$ are resonant, and the feed is effectively a pair of
parallel half-wave dipoles separated by about $0.45{\lambda}$ and close to
$0.15{\lambda}$ above the ground plane \citep{olss06}. We refined this basic
design using the CST simulation package \citep{cst14}. Starting from an array of
13 dipoles and a replication factor of 1.15, we used the optimization routines
within CST to improve the performance. The challenge in designing such a feed is
to maintain constant beamwidth and gain across the operating band while at the
same time achieving an acceptable impedance match. Using its internal
genetic-algorithm optimizer, we allowed CST to vary dipole length, the width of
the dipole arms, and the angle of the petal above the ground plane (the latter
parameter should be constant in a truly frequency-independent feed). Each
iteration was evaluated by considering beamwidth at the $-$10 dB points
(desirable width ${\sim}120^{\circ}$), the closely related forward gain
(desirable value 10 dB) and input return loss (desirable value 10~dB). Return
loss was based on a characteristic impedance of 200\,ohms for the balanced line.
A petal thickness of 3.2\,mm was assumed throughout the simulation (a departure
from log periodicity, more significant at the high end of the band). In the
final design the shortest and longest dipoles have lengths 9\,cm and 61\,cm
respectively, corresponding to frequencies of 1665 and 245 MHz respectively.

Each petal was connected by a twin-wire line to a balun. A Marchand balun was
used and impedance transformation, from 200 ohms on the balanced side to 50~ohms
on the unbalanced side, was built into it. The design is derived from the ideas
presented in \citet{pugl02}. A return loss $>10$\,dB was achieved across the
band 300 to 1300\,MHz; phase balance was within $5^{\circ}$.

\begin{figure}
\centering
\includegraphics[width=0.9\columnwidth,bb= 30 90 650 640]{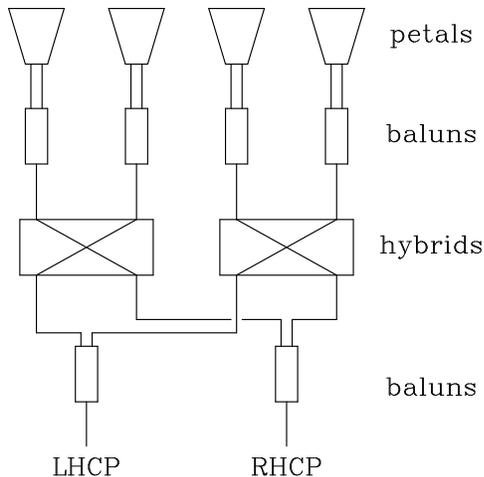}
 \caption{Diagram of the phasing network. The petals are the radiating elements,
connected through baluns to the $90^{\circ}$ hybrids, which combine signals in
quadrature. Two hands of circular polarization emerge from the lower layer of
baluns, which are used as $180^{\circ}$ hybrids.}
 \label{balun}
\end{figure}

Two sets of petals were placed orthogonally to accept two linear
polarizations. The petals were connected to a network that excited them at
phases $0^{\circ}$, $90^{\circ}$, $180^{\circ}$, and $270^{\circ}$ so that $L$
and $R$ could be generated: the phasing network is illustrated in
Figure~\ref{balun}. It employed TEM-line hybrids to generate $90^{\circ}$ phase
shifts and balun circuits to generate $180^{\circ}$ phase shifts. The baluns
were similar to those connected to the petals, but transformed between 50-ohms
unbalanced and two 50-ohm unbalanced  outputs in antiphase, the equivalent of
100\,ohms balanced. They were fitted with connectors for simple interconnection
of the phasing elements.

\subsection{Feed Fabrication and Construction}
\label{construct}

The petals were fabricated by water-jet cutting from 1/8 inch brass sheet. The
thickness was chosen because thick radiating structures have lower loss than
thin ones. The material was chosen to allow the baluns to be soldered to the
radiating elements. Each petal was glued to low-loss dielectric
foam{\footnote{Cuming Microwave C-STOCK RH5}} and the assembly was supported
above the ground plane on eight polystyrene rods 19\,mm in diameter. The
twin-wire line between the petals and the baluns was made of stranded copper
wire to maintain flexibility of this important connection (the feed is subject
to considerable vibration on the telescope). The baluns were fabricated as
printed circuits on standard FR4 circuit board. The TEM hybrids were commercial
devices{\footnote{R{\&}D Microwaves model HD-A01}}.

\subsection{Feed Performance}
\label{feedperform}

\begin{figure}
\centering
\includegraphics[width=0.9\columnwidth,bb= 10 20 600 610]{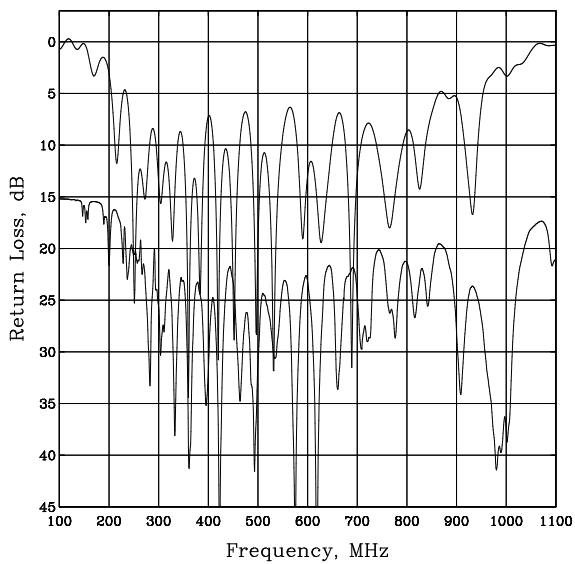}
 \caption{Calculated return loss of one polarization of the feed (top curve).
Measured return loss of one petal (bottom curve - displaced by 15\,dB for clarity).}
 \label{returnloss}
\end{figure}

Figure~\ref{returnloss} presents results for the return loss of a single petal
correctly placed above the ground plane. Measurements were made through the
balun. The other three petals were terminated with matched resistors. Simulation
suggested that the return loss of the feed as a whole closely resembled that of
a single petal, but the return loss of the assembled feed was impossible to
measure directly. It is clear that the manufactured feed matched the simulation
well, which gave confidence in our ability to simulate the feed and demonstrated
that the balun is essentially transparent. However, we see that our design did
not reach its target; the worst-case return loss was $\sim$5\,dB, and in narrow
frequency ranges about 30\% of the power in the incoming signal was reflected.
In the band for which results are presented in this paper, 300 to 480\,MHz,
minimum return loss was 7\,dB. Averaged over that band, the net power transfer
efficiency to a perfectly matched termination would have been 93.4\%, but
we note that poor matching of the feed was confined to narrow and separated
frequency ranges. The receiver components, especially the Low-Noise Amplifier,
were not perfectly matched, so power transfer may have been slightly higher or
lower than this figure. No correction was necessary for feed mismatch because the primary calibrator, Taurus\,A, (see Section~\ref{calib}) was, of course, observed through the same feed.

\begin{figure}
\centering
\includegraphics[width=0.9\columnwidth,bb= 20 10 580 700]{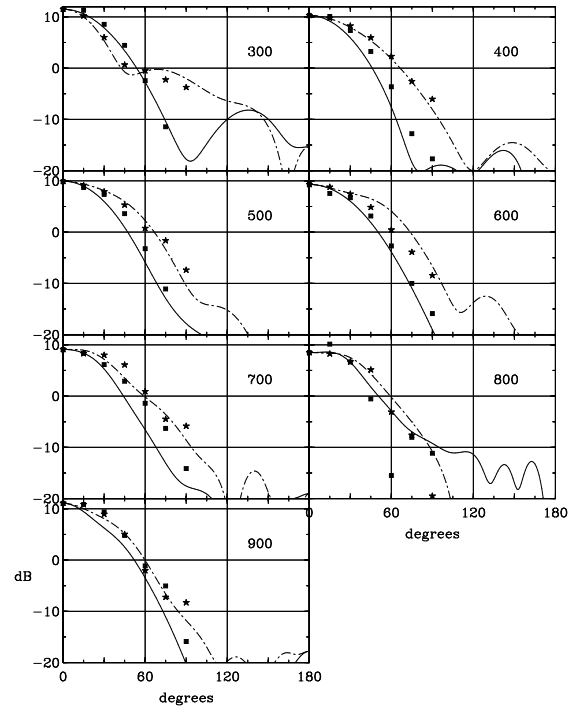}
 \caption{Radiation patterns of the feed. Calculated patterns: E-plane - solid
lines, H-plane - dash-dot lines. Measured patterns: E-plane - squares, H-plane -
stars. Frequency (MHz) is shown in each panel.}
 \label{radpattern}
\end{figure}

Measuring the feed radiation characteristics was more challenging: the feed
was big and heavy, and we did not have access to a sophisticated antenna range
where accurate measurements would have been possible. We simply set up the feed
outdoors, about one metre above the ground, and made measurements on a circle of
radius 10 m centered on the feed. We could measure E- and H-plane patterns
approximately, but it was not possible to measure cross-polarization
performance. Figure~\ref{radpattern} shows the resulting radiation pattern
measurements together with patterns calculated in the simulations. At all
frequencies below 700\,MHz the feed radiation patterns were symmetrical
(although this is not shown in Figure~\ref{radpattern}) and the patterns were
approximately circular. We judged that radiation past the edge of the reflector
(spillover) would be at an acceptably low level.

Measured feed performance at 700 and 800\,MHz departed from simulations. This may be connected with an abrupt change in measured beamwidth of the telescope at $\sim$800\,MHz - see Section~\ref{calib} and Figure~\ref{beam}. 

\subsection{Losses in the Feed and Phasing Network}
\label{losses}

Losses in the feed and the phasing network affect the calibration of the survey.
The insertion losses of these components were measured directly wherever
possible using a network analyzer. Where direct measurement was not possible,
measurements of return loss were used (also made with a network analyzer). When
the output of a device is short circuited a signal injected into its input will
pass through the device, will be reflected, and will return to the input. The
signal passes through the device twice, and the insertion loss is then half the
measured return loss.

To measure the loss in the feed a signal was transmitted into it and reflected
back into its input. To achieve total reflection, the feed was completely
surrounded by a metal box $1.2 \times 1.2 \times 1.04$\, m. This box contacted
the ground plane of the feed (size $1.2 \times 1.2$\,m). Two measurements were
made, one with a box depth of 1.04\,m and one with a depth of 0.9\,m; the
results were identical within the errors. This technique is essentially the
Wheeler cap method \citep{whee59}.

The 200-ohm baluns were soldered in place and could not be measured separately.
It was assumed that they have the same loss as the 50-ohm baluns (a reasonable
assumption since the loss was mostly in the FR4 circuit board).

\begin{figure}
\centering
\includegraphics[width=0.9\columnwidth,bb= 40 80 610 670]{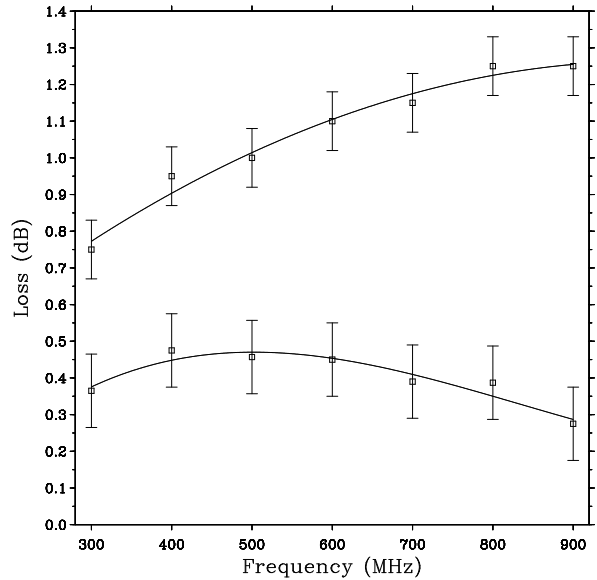}
\caption{Measured loss in the feed and circuits that precede the point where the
calibration noise signal is injected. The upper curve displays the total
measured loss, and the lower curve the loss in the feed.}
\label{feed_loss}
\end{figure}

Total losses in the feed and phasing network range from 0.80~dB at 300~MHz to
1.1~dB at 500~MHz and 1.26~dB at 900~MHz. Loss in the feed itself was calculated
by subtracting balun loss from the results of the measurement described above.
The estimated accuracy of the losses determined in this series of measurements
is $\pm 0.25$ dB or $\pm 5$\%. Some error is expected because currents in a
device with its output short-circuited are not indentical with currents under
normal use, but the technique certainly gives credible upper limits to loss.
Measured losses are summarized in Figure~\ref{feed_loss}.  

We note that the Eleven Feed has fairly high loss. The signal at any frequency
in the band travels from the dipole actively receiving that frequency through
all the shorter dipoles before reaching the balun.  In our implementation the
length of that path is about one wavelength at any given frequency in the band.

\subsection{Receiver}

The receiver was very straightforward. It had no frequency conversion, but
simply amplified the received signals to a level where they could be transmitted
via coaxial cable from the telescope focus cabin to the receiver room (a
distance of about 100 m).  The low-noise amplifiers (LNAs) employed were
commercial devices which gave a very flat passband but not exceptionally low
noise temperature. The excess noise, $T_e$, of the LNAs was $\sim$70~K. System
temperature varied from 115~K at 300\,MHz to 145~K at 900\,MHz without the feed.
Losses of 0.8 to 1.2\,dB in the feed and phasing network (see
Figure~\ref{feed_loss}) added 55 to 85\,K to the system noise.

Bandpass filters following the LNAs defined two bands, 290 to 470\,MHz and 660
to 870\,MHz. Outside these two bands the level of RFI was so high that radio
astronomy measurements were not possible.  Received signals were analyzed with a
digital spectropolarimeter of input bandwidth 1\,GHz, able to form all products
$LL^*, RR^*, RL^*$ and $LR^*$ from the input $L$ and $R$ signals, allowing
the derivation of all four Stokes parameters in 2048 frequency channels. The
Stokes parameters were calculated as ${I}={0.5{\thinspace}(RR^*+LL^*)}$,
${Q}={LR^*}$, and ${U}={RL^*}$. A noise source in a temperature-controlled
enclosure was coupled equally into both $L$ and $R$ paths with equal phase
(after the feed and phasing network) providing a linearly polarized calibration
signal.

\section{Survey Observations}
\label{obs}

\begin{table*}[!ht] 
\caption{Survey parameters} 
\label{params} 
\begin{center}
\begin{tabular}{@{}ll} 
\hline 
Beam FWHM & $81'$ at 300\,MHz, $45'$ at 480\,MHz\\ 
Channel bandwidth & 0.5\,MHz\\ 
Aperture efficiency & $\sim$27\%\\ 
Beam efficiency & 55\% at 300\,MHz, 43\% at 480\,MHz\\
Integration time & 0.25s\\ 
Declination range & -90$^{\circ}$ to +20$^{\circ}$\\
System temperature & 250\,K\\ 
Observing dates & 07-Sep-2009 -- 21-Sep-2009\\ 
 & 30-Nov-2009 -- 09-Dec-2009\\ 
 & 23-Feb-2010 -- 09-Mar-2010\\ 
 & 25-Jun-2010 -- 08-Jul-2010\\ 
 & 26-Aug-2010 -- 10-Sep-2010\\ 
 & 10-Nov-2010 -- 24-Nov-2010\\ 
 & 09-Feb-2011 -- 23-Feb-2011\\ 
 & 20-Oct-2011 -- 10-Nov-2011\\
 & 08-Feb-2012 -- 29-Feb-2012\\
 & 08-Jun-2012 -- 02-Jul-2012\\ 
Resolution in Faraday depth & 5.9\,${\rm{rad}}\thinspace{\rm{m}}^{-2}$ \\ 
Largest scale in Faraday depth & 8.6\,${\rm{rad}}\thinspace{\rm{m}}^{-2}$\\ 
Maximum observable Faraday depth & 1700\,${\rm{rad}}\thinspace{\rm{m}}^{-2}$\\ 
\hline 
\end{tabular} 
\end{center}
\end{table*}

All survey observations used the technique developed for the S-band Parkes
All-Sky Survey with the same telescope \citep{carr19}. The telescope was scanned
rapidly, at $15^{\circ}$ per minute, in azimuth, $A$, at the elevation of the
South Celestial Pole, $33^{\circ}$, with the feed fixed in the telescope frame.
Ground radiation is essentially constant as a function of azimuth (the terrain
around the telescope is quite flat). Fixed contributions to the observed $Q$ and
$U$, particularly ground radiation and instrumental polarization, remained
constant throughout a scan. The polarized signal from the sky, however, was
modulated by the changing parallactic angle, producing a sinusoidal variation
along the scan. The two contributions could therefore be separated in data
processing, and the ground and instrumental contributions removed, preserving
the sky signal on all scales. The key to successful application of this
technique is long azimuth scans, to produce the greatest possible range of
parallactic angle.

Scans were either ``East'' scans (${A}<{180^{\circ}}$) or ``West'' scans
(${A}>{180^{\circ}}$), where ${A}={180^{\circ}}$ is South. We observed either
the setting sky or the rising sky: this strategy produced scans that crossed one
another at large angles, enhancing the benefits of basketweaving (see
Section~\ref{basket}). West scans ran from $A=180^{\circ}$ to $A=290^{\circ}$.
Short East scans ran from $A=180^{\circ}$ to $A=30^{\circ}$ and long East scans
ran from $A=180^{\circ}$ through $0^{\circ}$ to $A=290^{\circ}$. This scanning
potentially gave complete coverage of the sky from declination
${\delta}={+24}^{\circ}$ to ${\delta}={-90}^{\circ}$. However, the master
equatorial, to which telescope movement is locked, could not be driven further
south than ${\delta}={{-87}^{\circ}}$.  West scans covered declinations
$-87^{\circ}$ to $0^{\circ}$ and East scans covered declinations $-87^{\circ}$
to $+20^{\circ}$. Data for the small area around the South Celestial Pole was
acquired during ramp-up time for scans nominally starting at  
${\delta}={{-87}^{\circ}}$.
Start times of scans were not arbitrary, but were spaced by
$10.5'$, chosen to ensure complete (Nyquist) sampling of the sky at 900\,MHz,
where the half-power beamwidth is ${\sim}25'$.

Observations were made exclusively at night to avoid contamination from solar
emission entering through sidelobes. Observations were scheduled in blocks of 12
to 20 nights, spaced throughout the year so that all Right Ascensions could be
covered. Altogether about 2000 hours of observing time were allocated, on the
dates listed in Table~\ref{params}.

\section{Absolute Calibration}
\label{calib}

We have made it a priority that all the surveys that comprise the GMIMS dataset
should be absolutely calibrated. With their wide frequency coverage, these
surveys extend far beyond the traditional radio astronomy frequency allocations,
and few or no absolutely calibrated data are available that we can tie our
surveys to. While the existing absolute calibrations are technically very
strong, they are old, and it is worthwhile to corroborate them with modern
measurements. 

In the calibration of a polarization survey there are two separate challenges:
calibration of the brightness temperature scale and calibration of polarization
angle. We were able to calibrate the brightness temperature scale with simple
techniques, but we were not able to calibrate polarization angle and eventually
had to rely on published data for the latter (see Section~\ref{anglecal}). There
are strong sources with high fractional polarization ({\it{e.g.}} 3C\,286,
3C\,138) whose characteristics are very well known; they are useful calibrators
of  polarization angle for telescopes of high angular resolution at frequencies
above $\sim$1\,GHz. At the frequencies that we used in this project, 300 to
900\,MHz, there are no strong polarized sources in the Southern sky, certainly
none strong enough to yield a significant signal in the beam,  $0.4^{\circ} -
1.3^{\circ}$. In this frequency range it is possible to calibrate using pulsars
({\it{e.g.}} \citealp{liao16}) but that option was not open to us.

The goal of absolute calibration is to measure the sky brightness temperature
in Kelvins on a scale tied to absolute standards of thermal noise, resistors at known temperatures. To achieve this we need to measure or calculate some telescope characteristics. 

We define the radio telescope by its power response $f(\theta,\phi)$, where $\theta,\phi$ are spherical coordinates and $f(0,0) = 1$. If the telescope is immersed in a temperature distribution $T(\theta,\phi)$ the available power at its terminals is equivalent to a temperature, the antenna temperature,
\begin{equation}
\label{antemp}
{T_A} = {\frac{1}{\Omega}}\thinspace{\int_{4\pi}}{T(\theta,\phi)}{f(\theta,\phi)}
{d{\omega}}.
\end{equation}
Here $d\omega$ is the element of solid angle, and $\Omega$ is the antenna solid angle defined by
\begin{equation}
\label{omega}
{\Omega}={\int_{4\pi}}{f(\theta,\phi}){d{\omega}}.
\end{equation}
A large reflector antenna like the Parkes telescope directs feed radiation into a main beam and sidelobes. Choosing a convenient boundary between them, we can separate $\Omega$ into main-beam solid angle, ${\Omega}_B$, and sidelobe solid angle, ${\Omega}_S$, and we can similarly separate $T_A$ into main beam and sidelobe contributions, $T_{AB}$ and $T_{AS}$. The closest approximation to the true brightness distribution over the main beam that the finite aperture of the telescope allows us to measure is
\begin{equation}
{T_B}={\frac{\Omega}{{\Omega}_B}}{T_{AB}}={\frac{\Omega}{{\Omega}_B}}
({T_A}-{T_{AS}}).
\end{equation}
Derivation of $T_B$ therefore requires correction of observations for the
contribution from sidelobes, which in turn demands full knowledge of the
sidelobes. This is a difficult assignment when mapping the total-intensity sky
with a large telescope. The problem is circumvented for the polarized sky
because $Q$ and $U$ take on positive and negative values and tend to average to
a very small number over large areas. Exceptions arise for spillover lobes
interacting with the ground: spillover lobes, far from the telescope axis, have
very non-ideal polarization properties, and can convert unpolarized ground
emission to apparently polarized signal. Our mapping technique overcame this,
and ground signal was effectively removed from the signal stream (see
Section~\ref{basket}).

The antenna solid angle, $\Omega$, is a measure of the ability of the telescope to concentrate radiation from the feed into a beam, or conversely, the ability of the telescope to collect radiation from a small area of the sky. This ability can (equivalently) be expressed in terms of an effective area of the telescope, $A_e$. Contrasting the antenna with an isotropic radiator (a theoretical construct) the gain of the antenna over the isotropic radiator is 
\begin{equation}
\label{asube}
G ={\frac{4\pi}{\Omega}}={\frac{{4\pi}{A_e}}{\lambda^2}}.
\end{equation}
The aperture efficiency, $\eta_A$ describes the utilization of the telescope aperture: it is the effective area, $A_e$, divided by the physical area, $A_p$,
\begin{equation}
\label{apeff}
{\eta_A}={\frac{A_e}{A_p}}.
\end{equation}
The beam efficiency is
\begin{equation}
\label{beameff}
{\eta_B}={\frac{\Omega_B}{\Omega}}.
\end{equation}
 
Several approaches to absolute calibration are possible. $\Omega$ can be measured by measuring the power response, $f(\theta,\phi)$ for all directions $(\theta,\phi)$. This is impractical for the Parkes telescope. $\Omega$ can be calculated from the known dimensions of the telescope, but \citet{du16} show that simulation software cannot yet achieve the desired precision. Finally, the increase in antenna temperature, ${\Delta}{T_A}$, when a radio source of known flux density $S$ is centered in the main beam can yield $\Omega$ through the relationship
\begin{equation}
{{\Delta}{T_A}}= {\frac{S\thinspace{\lambda}^2}{2\thinspace{k}\thinspace{\Omega}}}.
\end{equation}
Here $k$ is Boltzmann's constant and flux density, $S$ is measured in Jy. We chose this method because there are strong calibration sources that have well-established flux densities, accurate to a few percent.

There are no strong, compact sources in the
Southern sky with precisely known flux densities and spectral indices, so we
chose one from the Northern sky. The strong Northern sources with accurately
known flux densities that are within the declination range of the Parkes
Telescope are Virgo\,A (declination ${\sim}12.4^{\circ}$) and Taurus\,A
(declination ${\sim}22.0^{\circ}$). We chose Taurus\,A because it is the more
powerful of the two, with the better determined absolute spectrum. We discuss the value that we have adopted for the flux density of Taurus\,A in Section~\ref{apeff}.

\subsection{Measuring the Calibration Signal}
\label{equivalenttemperature}

The equivalent temperature in Kelvins of the injected noise signal was measured
in June 2010, February 2011, and February 2012. Measurements were made
separately in the $L$ and $R$ channels relative to resistive terminations at
ambient temperature and in a liquid-nitrogen bath. These measurements were
straightforward. Calibration signal amplitude was 24 to 47\,K across the
frequency band. The error, estimated from differences in the three measurements,
was $\sim$0.7\,K. Feed mismatch, discussed in Section~\ref{feedperform}, had no effect on the calibration signal, or on the accuracy of our measurements of it. The calibration noise signal always worked into a well matched termination, and only a small fraction of the noise power was coupled into the reciever through directional couplers.

\subsection{Measuring Aperture Efficiency}
\label{apeff}

\begin{figure}
\centering
\includegraphics[width=1.05\columnwidth,bb= 50 80 620 500]{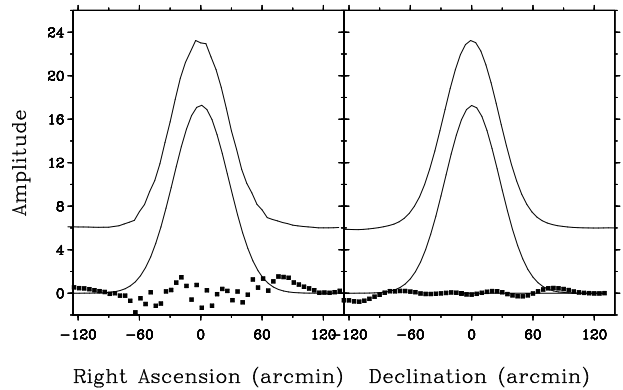}
 \caption{Cross-sections in Right Ascension and Declination through the telescope beam at 365.75\,MHz. The amplitude units are arbitrary. The top curves, displaced vertically by 6 units, show the measured data. The lower curves are Gaussians, fitted to the data. The dotted curves show residuals, multiplied by 5 for clarity. Scanning was in the Declination direction, and some scan-to-scan variations appear as additional residuals in the Right Ascension scan.}
 \label{gfit}
\end{figure}

At least once in every observing session (see Table~\ref{params}) a raster
scan was made of Taurus\,A covering an area ${5^{\circ}}\times{5^{\circ}}$. The
beamshapes deduced from these observations were very closely Gaussian in
profile. We averaged ten channels between 360.75 and 370.75\,MHz (near the
centre of our survey band). We removed a twisted-plane baseline from the map
(justified since Taurus\,A is near the Galactic plane). In Figure~\ref{gfit} we
show cross-sections through the beam in Right Ascension and Declination and
Gaussian fits to those profiles. Although the Gaussians were fitted to the
individual cross-sections, they are consistent in width within the errors. $LL$
and $RR$ beams were coincident.

\begin{figure}
\centering
\includegraphics[bb= 40 80 610 600,width=0.9\columnwidth]{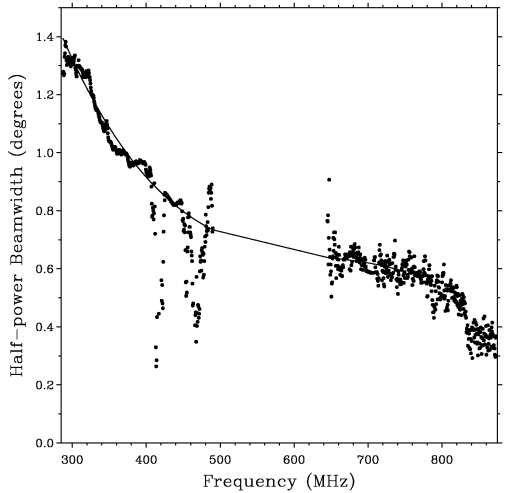}
 \caption{The beamwidth of the telescope, measured from raster scans of Taurus
 A. Twelve separate scans made on different days throughout the survey period
 were averaged to produce the data in this figure. The presence of RFI is 
 obvious. Fluctuations are higher above 660\,MHz, probably due to RFI,
 and no data from that band were used in this calibration. A cubic 
 polynomial, fitted to the data up to 750\,MHz is shown; this curve was used in
 absolute calibration of the survey. The rapid fall in deduced beamwidth above 
 $\sim$800\,MHz is not completely understood (see text).}
 \label{beam}
\end{figure}

The individual raster scans of Taurus\,A were processed as follows. A
two-dimensional Gaussian over a twisted-plane base was fitted to the map and the
amplitude and half-power widths of the Gaussian were tabulated. Widths were
corrected for the finite extent of Taurus\,A. Figure~\ref{beam} shows the
half-width of the Gaussian, averaged over twelve scans, and the polynomial
fitted to the average; telescope beamwidth was taken from this fitted curve.
Each of the twelve scans was affected by RFI. RFI-affected channels were flagged
and mostly did not participate in the average. Nevertheless, the impact of RFI
can be seen clearly in Figure~\ref{beam} from 400\,MHz to the top of the band
{\footnote{The rapid fall in deduced beamwidth above $\sim$800\,MHz is not
completely understood. It is possibly related to departures of measured feed
performance from simulations near that frequency discussed in
Section~\ref{feedperform}, but it may also be a result of remanent RFI in the
Taurus\,A data below or above 800\,MHz. We have not concerned ourselves with
this effect: this paper does not deal with data in that frequency range.}}.

The adopted flux density of Taurus\,A is 1450\,Jy at 300\,MHz with a
spectral index of $-$0.299. These values were taken from the VLSS Bright Source
Spectral Calulator (Ida10g.alliance.unm.edu/calspec/calspec.html) in 2008. The
flux density value was consistent with the spectrum for Taurus\,A established by
\citet{baar77}, which covered frequencies down to 1000\,MHz, but the website
entry also took into account more recent data at lower frequencies, and
therefore seemed to us to account for a possible decline in the flux density of
the source. However, the website flux density has subsequently been revised to
1407\,Jy, citing errors in the survey paper \citep{helm08}. This
notwithstanding, we believe 1450\,Jy is a good value for the following reasons.
Taking numbers from \citet{baar77} and extrapolating beyond 1000\,MHz to
300\,MHz gives a flux density of 1494\,Jy for epoch 1980. As might be expected
for a supernova remnant, the flux density of Taurus\,A is apparently declining:
\citet{alle85} measured a decline of 0.167\% per year at 8.0\,GHz and
\cite{viny07} find a similar rate of decline at 151.5 and 927\,MHz. Applying
this rate of decline to the \citet{baar77} value over the 30 years from 1980 to
2010 gives a flux density of 1419\,Jy at 300\,MHz. Our adopted value of 1450\,Jy
is within 3\% of all of these numbers. We accept a slightly higher number, 5\%,
as the probable error in our adopted flux density: this covers the range of
values permitted by the parameters for Taurus\,A given by \citet{baar77}. Our
adopted spectral index, $-0.299\pm0.009$, is the spectral index derived by
\citet{baar77}. We note that many of the measurements used by \citet{baar77}
were made with horn antennas and dipole arrays for which the conversion from
Kelvins of antenna temperature to Janskys can be accurately calculated. These
are therefore absolute measurements, and we consider that our survey is
absolutely calibrated.

\begin{figure}
\centering
\includegraphics[bb= 50 200 570 460,width=0.9\columnwidth]{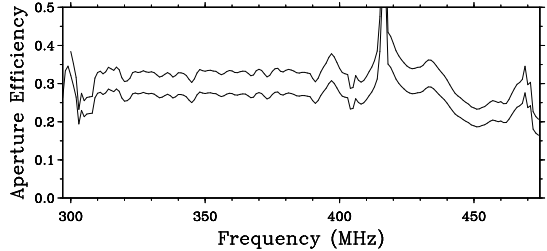}
 \caption{The lower curve shows the measured aperture efficiency of the 
 telescope. The best estimate of aperture efficiency is a constant value of
 ${{\eta}_A}={0.27{\pm}0.01}$ across the band - see text. The values shown here
 include losses in the feed and phasing  network. The upper curve shows the
 aperture efficiency corrected for those losses - see Figure~\ref{feed_loss}.}
 \label{ap_eff}
\end{figure}

The amplitudes of the fitted Gaussians were converted to temperature units by
comparison with the calibration noise source. This gave the antenna temperature,
$T_{A}({\rm{Tau A}})$, as a function of frequency. Given the adopted flux
density and spectral index, we were able to calculate the aperture efficiency,
${\eta}_A$, of the telescope;  the result is shown in Figure~\ref{ap_eff}.
Aperture efficiency measured in this way includes the loss of the feed and
phasing network (see Section~\ref{losses} and Figure~\ref{feed_loss}). A second
curve in Figure~\ref{ap_eff} shows aperture efficiency corrected for feed loss. 
It is very unlikely that $\eta_A$ has fine frequency structure. We believe that
the apparent depression in $\eta_A$ below 320\,MHz and the rapid fluctuations of
$\eta_A$ from 380 to 480\,MHz arise from vestiges of RFI in the data. The best
approximation to $\eta_A$ is a constant value of $0.32{\pm}0.01$ across the
band after correction for loss in the feed.

We expected a value of ${{\eta}_A}\approx{45}$\% for the Parkes Telescope
with a feed of this type. From the radiation patterns of the feed
(Figure~\ref{radpattern}) we calculated illumination efficiency, $\eta_I$, and
spillover efficiency, $\eta_S$; from telescope dimensions we calculated blockage
efficiency, $\eta_B$, taking into account both plane-wave and spherical-wave
blockage (see \citealp{du16} for an explanation of these terms). At our low
frequencies other factors (surface errors and surface transparency) are
negligible for the Parkes telescope. We derived ${\eta_I}\approx{0.55}$,
${\eta_S}\approx{0.85}$, and ${\eta_B}\approx{0.9}$, and their product gave a
predicted ${\eta_A}\approx{0.42}$. What can have reduced this value to 0.32? Our
hypothesis is that the feed was not quite at the focus of the telescope, but we
now have no way of confirming this.  A phase error introduced by a mis-placed
feed would have reduced gain and raised the sidelobe level in the forward
hemisphere, but would not have introduced sidelobes in any particular direction.
We note that the absolute calibration process can be relied upon to give the
correct answer, whatever aperture efficiency was achieved.

The error in our measurement of aperture efficiency, and consequently in the
intensity scale of our data, is probably of order 5\%. To this must be added a
further 5\% for the probable error in the adopted flux density of Taurus\,A, yielding an overall accuracy for the amplitude scale of 7\%.

Knowing the aperture efficiency allowed us to convert the survey data to
antenna temperature, $T_A$, but the desired product from the survey is
brightness temperature, $T_B$. The two are related by 
\begin{equation}
{T_B}={\frac{\Omega}{\Omega_B}}{\thinspace}{T_A}. 
\end{equation} 
Given that the telescope beams
are closely Gaussian, $\Omega_B$ was taken to be the solid angle of a Gaussian
of half-power width, $\theta$, equal to the measured beamwidth (see
Figure~\ref{beam}). Then ${\Omega_B}= 1.13{\thinspace}{\theta^2}$.

\section{Data Processing}
\label{process}

Raw data were recorded in the RPFITS format, with each file containing a single
azimuth scan or raster map. These files were converted into SDFITS, using the
program {\tt{rp2sdfits}} of the ATNF {\tt{livedata}} library, and finally
converted into our own binary format for further calibration and processing with
C++ code written especially for this survey. We will now discuss the particulars
of most of the data processing steps in the order applied.

\subsection{Primary Calibrators}

At least once a night one of two primary calibrators (Fornax-A and Pictor-A) was
observed, usually around sunset or sunrise, depending on the time of year. 
Observations were made as raster scans resulting in a rectangular map centered
on the source. Scan separation was $12'$, yielding greatly oversampled maps at
the low end of the band. A two-dimensional Gaussian function was fitted to each
of these maps, with the free parameters allowing for an elliptical and tilted
beam. The derived amplitude was corrected for atmospheric attenuation (a small
effect, $<0.1\%$). Both these sources were slightly resolved by our beams, but
that is not a concern here, since we were using only the amplitude of the fitted
Gaussian.

Using this database of Gaussian fits, an average flux density and temperature
spectral index was determined for the two calibrators (Fornax-A: $-2.5$,
Pictor-A: $-2.1$). This is by no means an absolute flux calibration; it was used
only to remove gain and phase drifts across time, to remove the instrumental
frequency response from the data, and to corect on-axis instrumental polarization.

\subsection{Noise Source Calibrations}

The noise source provided a polarized and stable signal. It was switched on
every 2 to 3 hours for a duration of 5 minutes, alternating between ON and OFF
with a frequency of 1\,Hz. The resulting OFF measurements were averaged and
subtracted from the average of the ON measurements, providing noise-source
temperatures for all four correlation outputs ($RR^*, LL^*, RL^*, LR^*$) at each
frequency channel. The time series of these measurements was smoothed by
averaging and discarding outliers, allowing us to detect and correct gain
variations of the receiver between calibrations.  All data were divided by the
smoothed series with interpolation between measurements, resulting in data now
calibrated in units of noise source temperature.

\subsection{Instrumental Polarization}

Two effects contribute to instrumental polarization, (a) cross-hand leakage
between $L$ and $R$ channels occurring in the feed, the phasing network, and
(perhaps) in the receiver, and (b) cross-polarization in the feed radiation
characteristics. We have corrected on-axis polarization, but have not dealt with
the much more complex problem of cross-polarization in the antenna response.

In the absence of instrumental polarization a scan across an unpolarized source
would produce a Gaussian-like profile in total intensity channels $LL^*$ and 
$RR^*$ and no signal in the two cross-hand channels $LR^*$ and $RL^*$. In fact
$LR^*$ and $RL^*$ showed complicated structure, a combination of effects (a) and
(b). Using an iterative process, we found factors $f_U$ and $f_Q$ for each
channel which minimized the response in $LR^*$ and $RL^*$. The ``unpolarized''
source used was Fornax\,A. (Although the source is a synchrotron emitter whose
emission must be polarized, when the entire source was averaged in our large
beams it became essentially unpolarized). The ratio at the beam center of $LR^*$
and $RL^*$ to total intensity was determined for every Fornax\,A observation.
These values were taken as the best estimate of the on-axis instrumental
polarization. Smoothing in time was applied to these values, since rapid
variations of instrumental polarization were not expected. The factors $f_U$ and
$f_Q$ were stable with time within ${\pm}10\%$. Instrumental polarization
artefacts remain in the data along the Galactic plane, where $I$ emission is
very strong (see Figures~\ref{moments} and \ref{fd}). Small polarization
artefacts, the classical ``four-leaf clover'' response, remain in the $Q$ and
$U$ data around strong compact sources at a level of a few percent, and there is
some variation from one such source to another. The portrayal of the low-level
extended emission, the objective of this survey, is not impaired by these very
localized blemishes.

\subsection{Ground Radiation}
\label{ground}

The observing technique, azimuth scanning, assumes that emission from the ground
is constant with azimuth. In the frequency range of this survey that assumption
may not be quite true. The far sidelobes of the telescope, including the
spillover sidelobes, have very strong instrumental polarization ({\it{e.g.}}
\citealp{du16}). They convert unpolarized signal from the ground (or from the
sky) into apparently polarized signal. In the range 300 to 480\,MHz the ground
is definitely not a black body, it is a partly polarized emitter \citep{du16},
and it will also reflect sky signal into the spillover sidelobes. We estimated
the ground contribution to the data by plotting all scans against azimuth and
masking out emission from the Galactic plane. We binned these data in steps of
$20^{\circ}$ in azimuth, found the median in each bin, and interpolated between
these points with a spline function. This was done independently for each
frequency channel, and the fitted functions were subtracted from the data. The
corrections were 0.2\,K or less.

\subsection{Radio Frequency Interference}
\label{rfi}

RFI posed a major threat to this survey. Locally generated RFI was minimized by
turning off fluorescent lights, computers and other equipment (unused during our
night-time observations) in all buildings around the telescope. A monitoring
system in another receiver in the focus cabin was also shut down every night.

\begin{figure}
\centering
\includegraphics[bb=50 100 600 460,width=1.0\columnwidth,clip=]{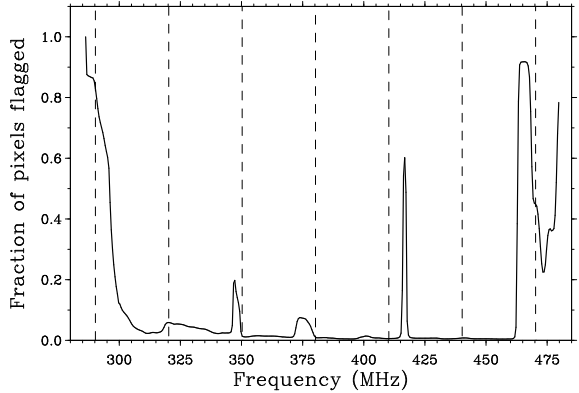}
 \caption{The fraction of the survey data flagged because of RFI, shown as a
 function of frequency. The vertical dashed lines indicate the frequencies used
 in the calculation of the T-T plots shown in Figure~\ref{internaltt}.}
 \label{flagging}
\end{figure}

The band 500\,MHz to 650\,MHz was totally occupied by digital television signals
broadcast from mountaintops 100 and 200\,km distant from the Observatory. The
band 490\,MHz to 660\,MHz was therefore completely  removed from the receiver
passband with filters, and no attempt was made to observe within it. The
receiver passband for the survey was determined by bandpass filters as 285 to
485\,MHz and 660 to 870\,MHz, which we refer to as the lower and upper bands
respectively. RFI, both steady and intermittent, occurred in both lower and
upper bands; in the upper band we lost 80\% of our observations to RFI, and in
the lower band we lost 50\%. Data loss in the upper band was so severe that we
have not yet been able to use the data in that frequency range. We have been
able to use data in the lower band because the sky was heavily oversampled
relative to the telescope beam in that band. The sampling interval was set at
$10.5'$, just under half the beamwidth at 900\,MHz. This sampling interval is
about one fifth of the beamwidth at 300\,MHz and one third of the beamwidth at
480\,MHz: at the low end of the band the sky was oversampled, and sky sampling
is still adequate even after removing RFI-affected data. We present data to a
lower frequency limit of $\sim$287\,MHz, but frequency channels below 300\,MHz
are heavily affected by RFI. 

Our RFI mitigation strategy depended heavily on oversampling in time and
frequency. The first step of RFI mitigation consisted of flagging outliers in a
time series. Every pixel was covered by many observations, spaced over 3.5
years, and RFI is time variable, especially over such a long period. For every
pixel we assembled a time series of observations, and calculated the median and
standard deviation for that series.  Every integration above or below the median
by about 3 standard deviations was flagged (flagging level varied a little with
frequency, and flagging level was set higher in regions of high sky brightness).
Median and standard deviation were calculated again, and more outliers flagged,
and the process was repeated until no integrations lay outside the thresholds.
The heavy spatial oversampling allowed us to set low thresholds for RFI
flagging.

The second step of RFI mitigation examined the data after processing to find
outliers along the frequency axis. A spectral index was determined for every
pixel and a baseline subtracted from the spectrum. All channels above or below a
certain threshold were flagged.  Figure~\ref{flagging} shows the fraction of
pixels flagged in the final data product. This number is low in most frequency
channels, attesting to the effectiveness of oversampling. That being said, the
final data products are not noise limited: they are probably limited by RFI at
even lower levels. 

\subsection{Ionospheric Faraday Rotation}

\begin{figure}
\centering
\includegraphics[bb= 60 60 570 530, width=0.65\columnwidth,clip=]{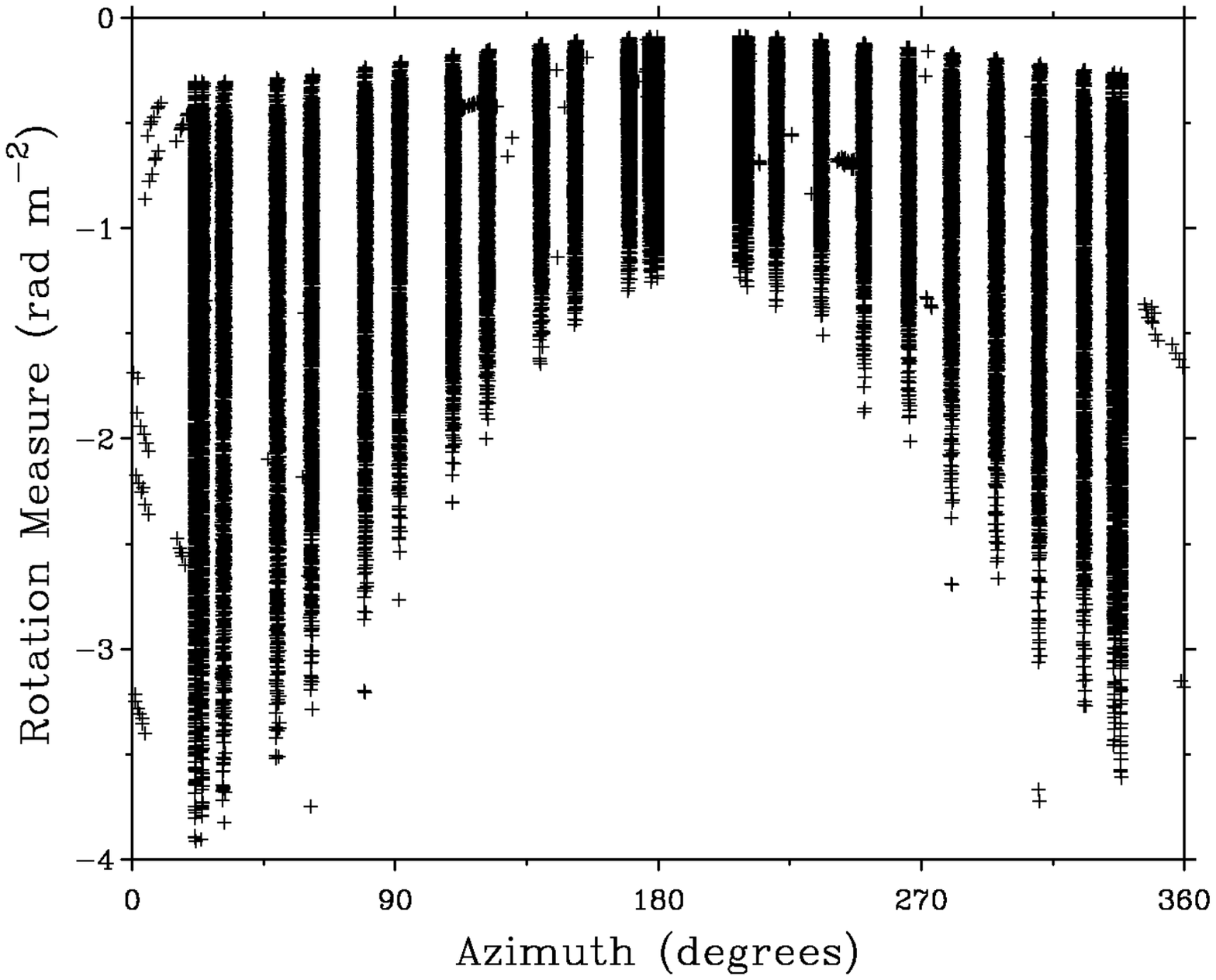}
\includegraphics[bb= 60 60 570 530, width=0.65\columnwidth,clip=]{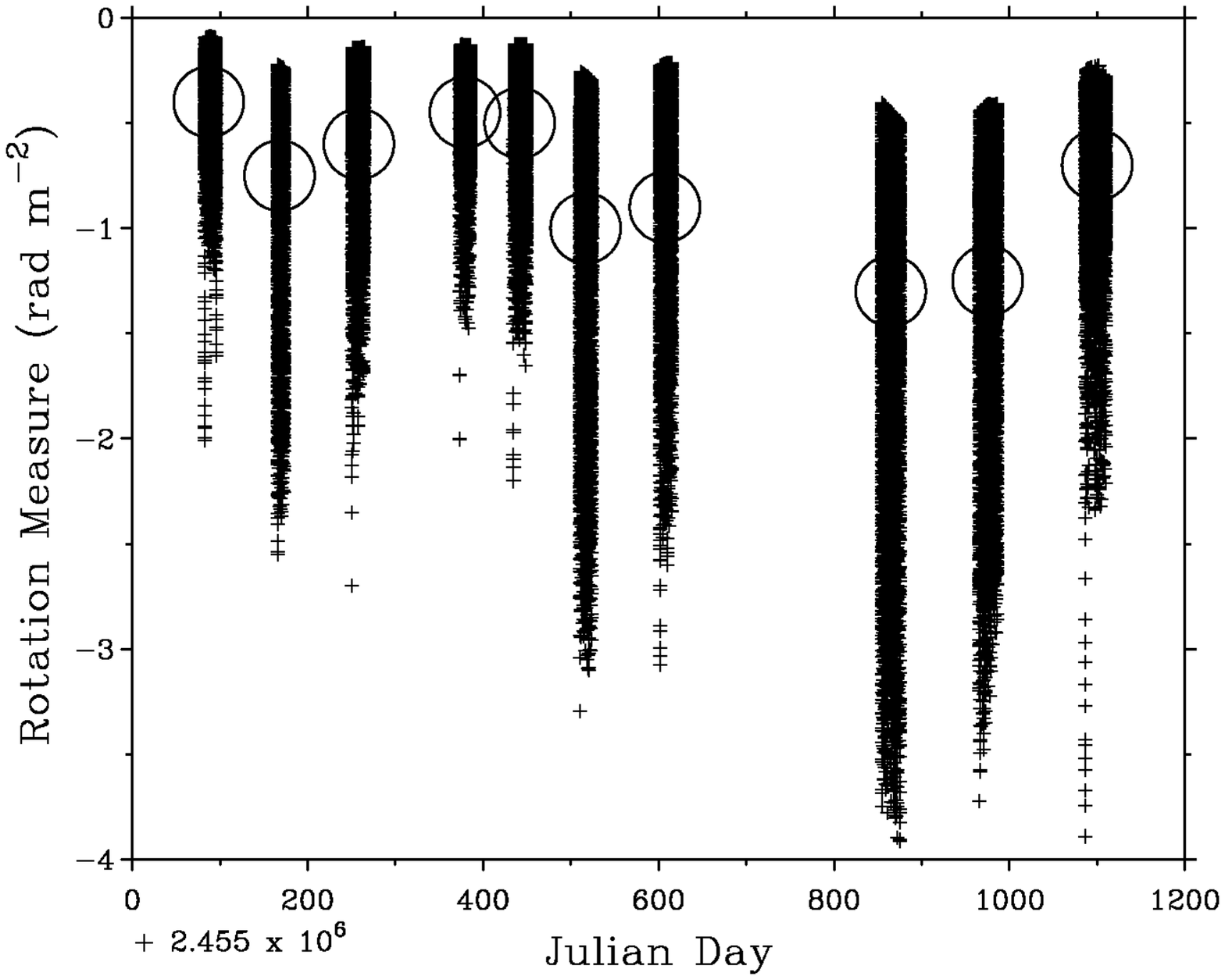}
 \caption{Top: predicted ionospheric Faraday rotation for all observations as a 
 function of azimuth. Faraday rotation was calculated at the azimuths shown and
 interpolated values were used for intermediate pointings. Bottom: ionospheric 
 Faraday rotation as a function of Julian Day. The ten blocks of data correspond
 to ten observing sessions (see Table~\ref{params}). Circles indicate median 
 values in individual observing sessions.}
 \label{iono}
\end{figure}

High electron densities in the ionosphere and a geomagnetic field component
along the line-of-sight give rise to Faraday rotation whose magnitude is highly
variable and dependent on solar activity. Observations for this survey were made
between sunset and sunrise, but spanned a period from solar minimum in 2009 to
one where the Sun was fairly active in mid-2012. The ionospheric RM ranged from
zero to about $-4\,{\rm{rad}}\thinspace{\rm{m}}^{-2}$. Since
$1\,{\rm{rad}}\thinspace{\rm{m}}^{-2}$ rotates the polarization angle by one
radian at 300\,MHz, this obviously had to be corrected.

Ionospheric Faraday rotation was corrected using an algorithm based on the
International Reference Ionosphere model (IRI - \citealp{bili15}) and a static
model of the geomagnetic field \citep{theb15}. The input quantity is the 10.7-cm
solar flux \citep{tapp13}. The model computes Faraday rotation through the
ionosphere as a function of direction. Figure~\ref{iono} shows the predicted
ionospheric Faraday rotation through the course of the survey.

More sophisticated routines are now available for calculation of ionospheric RM
({\it{e.g.}} \citealp{mevi18}) which use GPS data to calculate the total
electron content of the ionosphere. From the $\sim$50,000 corrections that we
computed (Figure~\ref{iono}) we selected 53 representative samples. We made new
calculations for those dates and telescope pointings using RMextract and using
the ALBUS code (A.G. Willis et al, in preparation 2019). The average ratio
between RMextract and our calculations was 1.155; the average ratio with ALBUS
was 1.005. We regard ALBUS as the superior code in the Australian context
because it uses many more GPS stations than RMextract. For $|{\rm{RM}}|<1.0$ the
peak difference between ALBUS and our calculations was
$0.2\,{\rm{rad}}\thinspace{\rm{m}}^{-2}$ and the rms difference was 0.07. For
$1.0 <|{\rm{RM}}|<3.0$ the peak difference was
$0.6\,{\rm{rad}}\thinspace{\rm{m}}^{-2}$ and the rms difference was 0.2. We
found one value where the GMIMS RM was $4\,{\rm{rad}}\thinspace{\rm{m}}^{-2}$
and ALBUS gave $5\,{\rm{rad}}\thinspace{\rm{m}}^{-2}$: this is the very highest
RM in the set of 50,000 calculations and we consider it an outlier.

In this survey, measurements taken at different times were combined
(vector averaged) using the basketweaving technique (see
Section~\ref{basket}). Errors in the ionospheric RM correction could have led to
a reduction in the measured polarized intensity below the true value, affecting,
for example, estimates of the fractional polarization (see Figure~\ref{fract}).
An error of $0.2\,{\rm{rad}}\thinspace{\rm{m}}^{-2}$ in the correction applied
to two polarization vectors that are then averaged leads to a reduction in
estimated polarized intensity of 2\% at 300\,MHz and correspondingly less at
higher frequencies. This is a small effect.

\subsection{Basketweaving}
\label{basket}

The observing technique, making many intersecting scans, has been described in
Section~\ref{obs}. Basketweaving, the reconciliation of intersecting scans, was
an integral part of the processing; our implementation mostly followed the
procedure of \citet{hasl81}.

The correlator output was the combination of several sources, receiver noise,
ground and spillover noise, and sky signal. While the sky signal changes with
position, the other contributions should remain constant. In reality, however,
any one of the system noise contributions may vary over time, caused by gain
variations of the receiver or by actual, intrinsic variations of the signal
({\it{e.g.}} variations of ground reflectivity with soil moisture). The
challenge in data processing was to separate the sky signal from the rest, and
to subtract that baseline with minimal effect on the sky signal.   The first
step in processing was to remove a linear baseline from each scan. The effect
was specific to the various correlator outputs. From the total intensity
products $LL^*$ and $RR^*$ this removed ground radiation, assumed constant
during the scan, and it removed system noise, likely to be highly stable over
the duration of one scan, but possibly variable on longer timescales; it also
removed some signal. From the polarization products $LR^*$ and $RL^*$ this
subtraction removed the ground radiation and the instrumental polarization
component, and the small component of receiver noise that coupled between $L$
and $R$ via polarization leakage, but removed very little, if any, sky signal.
The long scans in azimuth with the feed at a fixed orientation modulated the sky
polarization by the parallactic angle \citep{carr19} and the average of sky
signal in each $LR^*$ and $RL^*$ scan was always close to zero.: $LR^* (=Q)$ and
$RL^* (=U)$ tend to average to zero along long scans, especially at the low
frequencies of our survey. Parallactic angle rotation was then applied,
restoring very closely correct zero levels to $LR^*$ and $RL^*$.

Basketweaving was performed on every channel individually, and separately on all
four products, $LL^*$, $RR^*$, $LR^*$, and $RL^*$. Baseline fitting ran through
several loops. The first two loops fitted constants to each baselevel. For the
next few loops the differences were smoothed over ten degrees, and then for two
final loops over five degrees. We also detected low-level periodic variations.
We suspect these arose from gain variations of the amplifiers that were not
corrected by the noise source calibration. These baseline fluctuations were
removed by fitting sinusoids.

In the final analysis our data are not limited by thermal noise, but probably by
low-level RFI (Section~\ref{rfi}). The baseline fitting procedure that we
adopted removed some RFI. It could possibly have been taken further, but at the
risk of removing some real signal. Basketweaving did remove the sky minimum from
the $LL^*$ and $RR^*$ images, and any use of  the Stokes $I$ data product will
have to take that into account.

\subsection{Gridding and Smoothing}

To this point the data have remained in the form of scans, long azimuth tracks
across the equatorial grid. Maps in equatorial and Galactic co-ordinate frames
were made from the scans by a simple gridding process: scan values falling
within a square with sides one grid interval, centered on each equatorial grid
point, were averaged. Smoothing to the final angular resolution, $1.35^{\circ}$,
was done at this point.

\subsection{Rotation Measure Synthesis}
\label{rot_meas_synth}

As described by \citet{bren05} a survey of linear polarization that covers a
sufficiently broad range of frequency, or, more exactly, of wavelength squared
(${\lambda}^2$), can be inverted by a Fourier transform to Faraday depth space
($\phi$). Because of the many interfering signals that required flagging of some
or all data at specific frequencies, the Fourier transform routine employed in
this operation must of necessity handle missing data intelligently. Our starting
point was a cube of 360 $Q$ and $U$ images evenly spaced in frequency, spanning
300.25 to 479.75\,MHz in steps of 0.5\,MHz, each channel smoothed to the angular
resolution ($1.35^{\circ}$) at the lowest frequency. From these data we computed
Faraday depth cubes in $Q$ and $U$ covering
${-100}<{\phi}<{+100}\,{\rm{rad}}\thinspace{\rm{m}}^{-2}$ in steps of
0.5\,${\rm{rad}}\thinspace{\rm{m}}^{-2}$. Our RM synthesis
routine{\footnote{RM\_tools\_3D, version of January 31, 2018, available from
https://github.com/crpurcell/RM-tools.}} implemented the equations of
\citet{bren05}, programmed in {\tt{python}}. All frequencies were weighted
equally. 

The data cube emerging from this process represents sky emission convolved with
a Rotation Measure Spread Function (RMSF) that is approximately the Fourier
transform of the sampling in ${\lambda}^2$. In the perfect world this sampling
would be complete and uniform, and deconvolution would be simple, but our heavy
flagging of RFI-affected data produced RMSFs that had strong sidelobes and
differed from one pixel to the next. RMSFs were therefore calculated separately
for each pixel. The `dirty' $\phi$ spectra were deconvolved with the RMCLEAN
routine \citep{heal09}. Loop gain was 0.1 and the process was limited to 60\,mK
(the survey rms) or 1000 iterations. The Faraday depth spectrum at each pixel
was restored using a Gaussian function fitted to the RMSF for that pixel.
Finally, residuals were added back. The parameters of the resulting data are
listed in Tables~\ref{params} and \ref{products}.

\begin{figure}
\centering
\includegraphics[width=1.05\columnwidth,clip=]{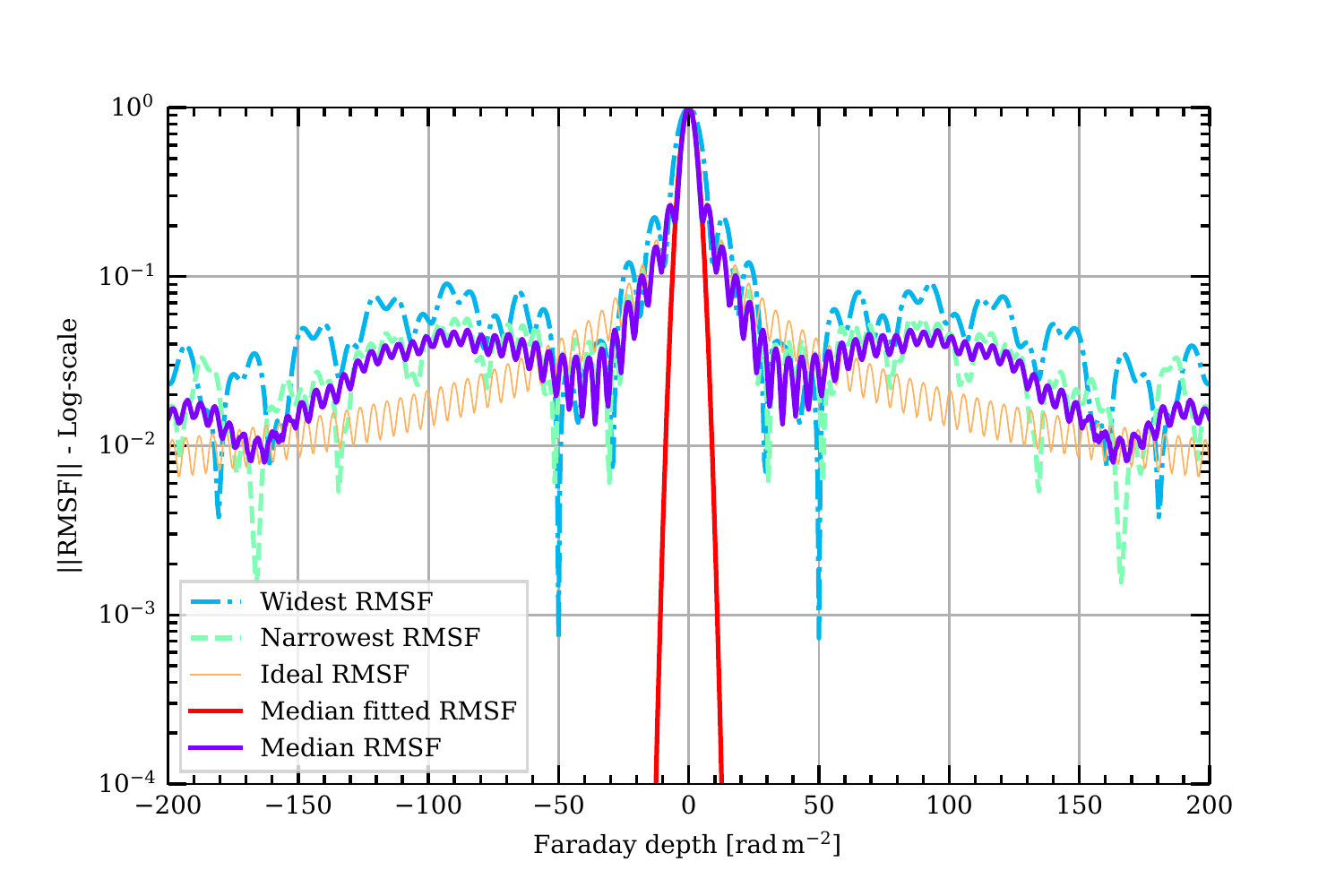}
 \caption{The Rotation Measure Spread Functions (RMSFs) of this dataset, shown over twice the range in Faraday depth of the final dataset. The narrowest, widest, and median RMSFs are shown, together with the ideal RMSF, as it would be if no frequency channels were flagged.}
 \label{rmsf}
\end{figure}

\begin{figure}
\centering
\includegraphics[width=1.08\columnwidth,clip=]{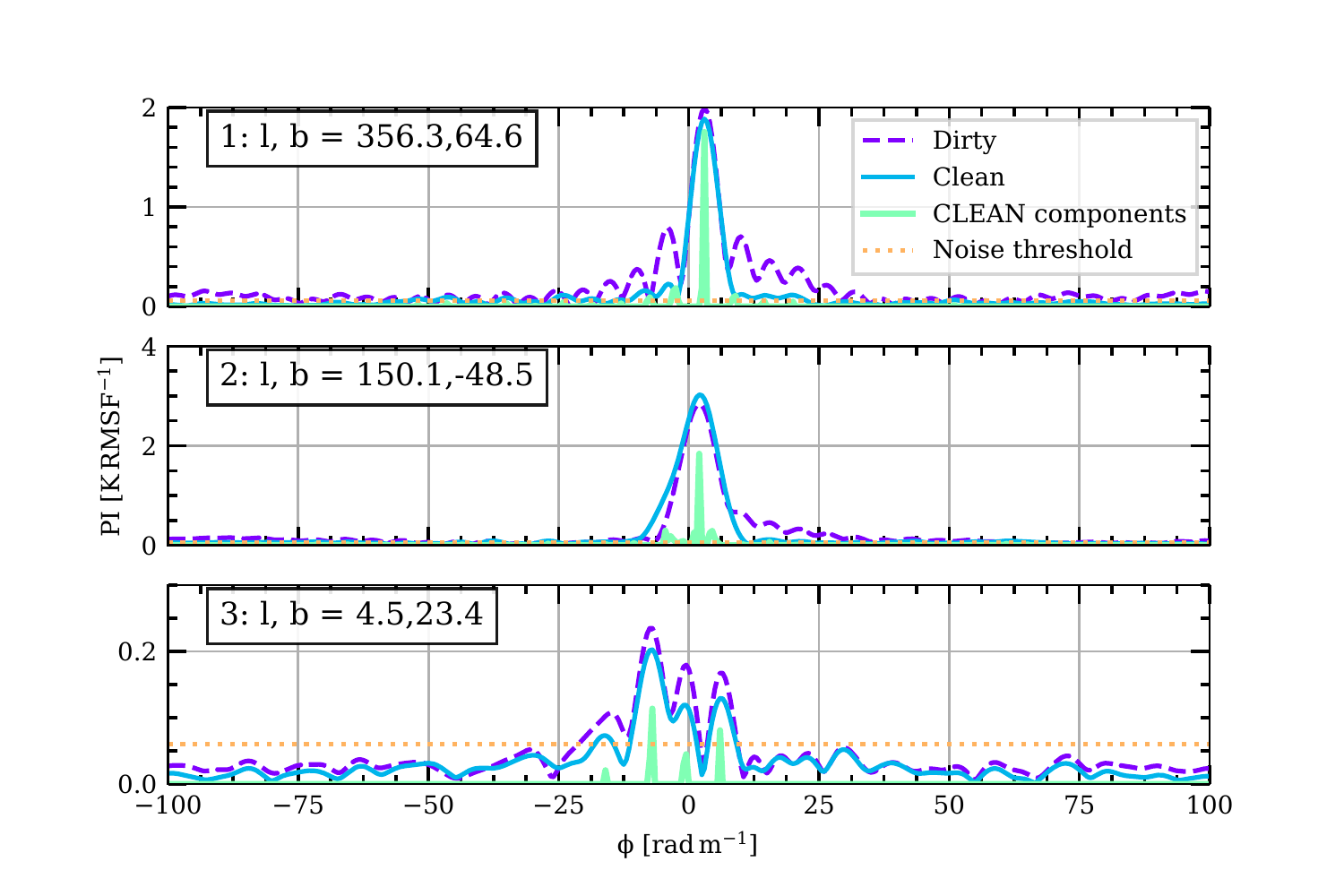}
 \caption{Faraday depth spectra for three lines of sight. The position in Galactic co-ordinates is indicated in each panel. See text for discussion of these spectra.}
 \label{cleandirty}
\end{figure}

Figure~\ref{rmsf} shows RMSFs relevant to this dataset.  The primary effect of
the flagged channels on the RMSF is to generate sidelobes at a level of
${\sim}4\%$ at ${\pm}90\,{\rm{rad}}\thinspace{\rm{m}}^{-2}$. The median RMSF
(median in width of the main lobe) is close to the narrowest RMSF, indicating
that flagging of RFI in the spectra retained in the dataset was quite modest
over much of the sky. Figure~\ref{cleandirty} shows both dirty and clean Faraday
spectra for three positions, each illustrating a different situation. Panel 1
shows a simple spectrum from an apparently Faraday-thin region. The dirty
spectrum has roughly symmetrical sidelobes on either side of the main peak; it
is satisfactorily fitted by a single clean component and the sidelobes are
almost completely removed by the cleaning process. Panel 2 shows a single
feature, broader than the RMSF, with a long tail towards positive $\phi$. It is
striking that the clean spectrum shows a tail towards negative $\phi$ (and the
sidelobes are almost completely removed). Panel 3 shows a low-intensity complex
spectrum. Note that the relative heights of the three peaks in the spectrum are
changed by cleaning. However, the significance of this change should not be over-emphasized: such occurrences are a common manifestation of the complex RMCLEAN algorithm \citep{sun15b}.

\section{Quality of the Survey Data}
\label{data_qual}

In order to assess the quality of the survey data we made comparisons with
existing data to the extent possible. Although the data presented here were
calibrated independently, comparison with other data helps with estimation of
errors.

\subsection{The intensity scale}
\label{assess_int}

The first step was to test the calibration of the total-intensity (Stokes
parameter $I$) scale. We used the T-T plot technique \citep{cost60}, in which
the brightness temperatures, $T_B$, at sky directions in one dataset are plotted
against the brightness temperatures in the same directions from another dataset.
If two datasets at the same frequency are compared in this way, the slope of the
straight line fitted to the points gives the ratio of the two intensity scales.
If the two datasets are at different frequencies, the slope of such a plot gives
the temperature spectral index, $\beta$, defined by ${T_B}\propto{\nu^{\beta}}$,
in that frequency interval (affected by any errors in the intensity scales ).

\begin{figure}
\centering
\includegraphics[bb=1 1 551 501,width=0.9\columnwidth,clip=]{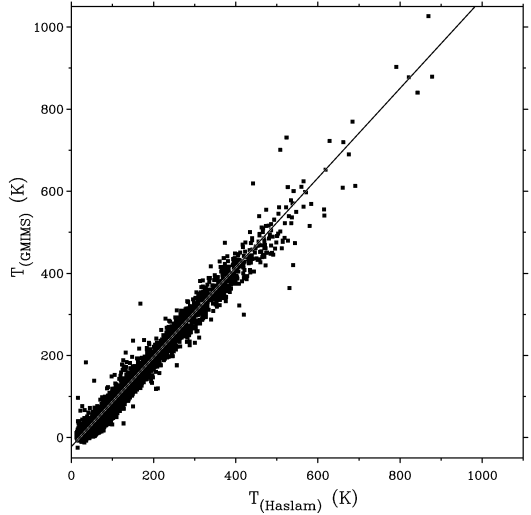}
 \caption{Brightness temperature at 408\,MHz plotted point-by-point against
 brightness temperature from the \citet{hasl82} survey at the same frequency.
 The plot includes data from the entire GMIMS survey area. Bandwidth in both 
 cases is 3.5\,MHz and beamwidths are $51'$. The fitted line has a slope of 
 1.09 and an offset of $-$19\,K.}
 \label{tt_408}
\end{figure}

We plotted total GMIMS intensity at 408\,MHz against corresponding values from
the survey of \citet{hasl82}. We used data at the resolution of the telescope,
$51'$ in both cases. Figure~\ref{tt_408} shows the result.  There are two
conclusions. First, the fitted line shown in Figure~\ref{tt_408} indicates that
the GMIMS intensity scale is 9\% higher than the \citet{hasl82} scale. That is a
satisfying result, given that the two surveys were independently calibrated to
absolute scales. Second, the T-T plot shows that the GMIMS zero level
is about 19\,K lower than the zero of the Haslam data. This is expected, because
the basketweaving process has removed the data minimum; the sky minimum in the Haslam data is about 13\,K (and the quoted error in the zero level is ${\pm}3$\,K).

We cannot say much about this 9\% difference, except that it is satisfyingly small. A direct comparison of our methods with the methods employed by \citet{hasl82} is difficult.  The \citet{hasl82} survey was calibrated by referring it to the earlier survey of \citet{paul62}, itself absolutely calibrated. The 1962 survey covered the sky North of declination $-20^{\circ}$, and the calibration factors established by \citet{hasl82} for the Northern sky had to be extrapolated to the South, creating an uncertainty (over most of the area of our survey) that we cannot evaluate. Furthermore, the \citet{hasl82} data were tied to the main-beam brightness temperatures from \citet{paul62}, eliminating steps of establishing main beam solid angle or beam efficiency for the 1982 data. Again, we cannot compare methods. Nevertheless, the \citet{hasl82} survey is well regarded, and its calibration is considered strong, and it is a satisfying outcome that the two intensity scales agree so well.

\begin{figure}
\centering
\includegraphics[width=0.9\columnwidth]{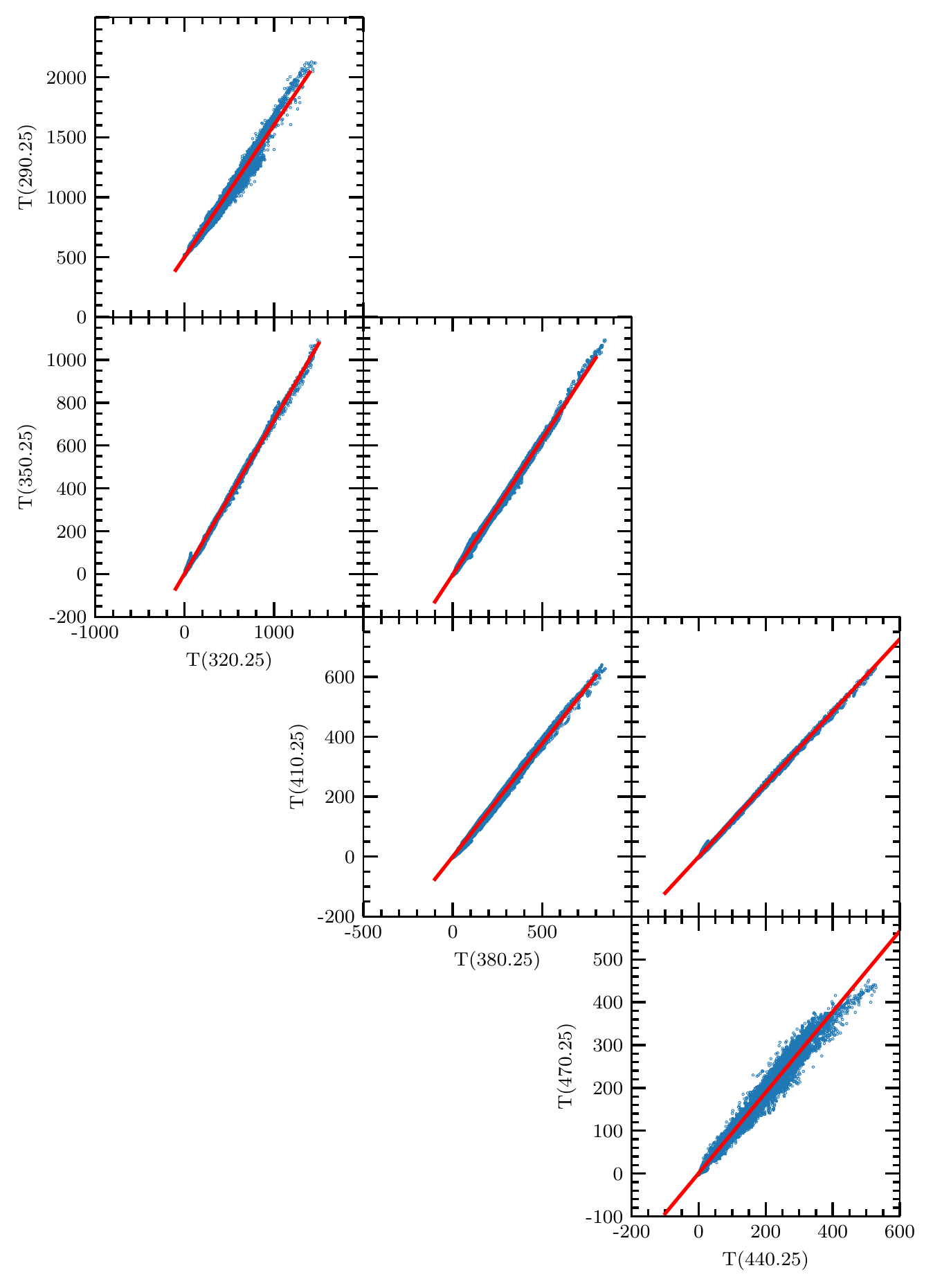}
 \caption{Spectral index of total intensity emission measured by plotting
 brightness temperature at one frequency against temperature at the
 corresponsing point measured at the second frequency. Plots are shown for
 six pairs of frequencies. See text for interpretation of these results.}
 \label{internaltt}
\end{figure}

Having satisfied ourselves that the intensity scale at 408\,MHz is correct, we
needed to check the scale at other frequencies across the band. We chose seven
frequencies from 290.25\,MHz to 470.25\,MHz at intervals of 30\,MHz (marked
in Figure~\ref{flagging}). At each of these frequencies we averaged five
channels, to give bands of width 2.5\,MHz. From these, we generated six T-T
plots, shown in Figure~\ref{internaltt}, between pairs of frequencies using all
the data points in the survey, that is over the entire Southern sky. The slope
of these T-T plots reflects the spectrum of the extended Galactic
emission. In this frequency range the emission is predominantly non-thermal,
with relatively little contribution from optically thin thermal gas.

If we know, or assume, a spectral index for the emission, we can check the
intensity scales at frequencies other than 408\,MHz. Data from  the Northern sky
indicate ${\beta}\approx{-2.5}$ between 151 and 408\,MHz \citep{siro74} and
${\beta}\approx{-2.8}$ between 408 and 1407\,MHz \citep{webs74}. Data for the
Southern sky between 408 and 720\,MHz give ${\beta}\approx{-2.8}$
\citep{land69}.

The correlations in Figure~\ref{internaltt} are tight: correlation coefficients
are above 0.994 for T-T plots from 320.25\,MHz to 440.25\,MHz. Correlation
coefficients of T-T plots involving the outer frequencies, 290.25\,MHz and
470.25\,MHz are not quite as high because there was considerable RFI in these
bands and many sky points were intentionally flagged (at 290.25\,MHz 86\% of
data was flagged, and at 470.25\,MHz 60\% was flagged).
The temperature spectral indices deduced
from the T-T plots in Figure~\ref{internaltt} vary from ${\sim}-0.9$ to
${\sim}-3.7$, but the frequency span for each calculation is very small, and
small inaccuracies in the temperature scales will produce large discrepancies in
deduced spectral indices.  We calculated correction factors for each of the six
T-T plots required to bring $\beta$ to a value of $-2.8$: these range from 0.994
to 1.178.  

However, this is not a very strong discriminator. If we choose any value of
$\beta$ between $-2.4$ and $-3.2$ for this frequency range, the correction
factors calculated as above are still in the range 0.9 to 1.1. All considered,
we conclude that the intensity scale across 300 to 480\,MHz is correct within
10\%. This calibration of total intensity immediately implies that polarized
intensities are correct to the same accuracy.

\subsection{Calibration of Polarization Angle}
\label{anglecal}

In an attempt to calibrate polarization angle, we interrupted observing at
two-hour intervals to make a rotating-feed observation of one of a number of
highly polarized regions. These regions were chosen on the basis of the data of
\citet{math65}, a 408\,MHz polarization survey made with the Parkes Telescope.
These calibrations were unsuccessful. We believe that distant sidelobes
passing over bright unpolarized regions when the feed was rotated generated
variations stronger than those arising from the polarized region at beam center.
Perhaps the feed used for this survey was not as highly tapered as the feed
used by \citet{math65}. See Section~\ref{ground} for a discussion of the
properties of distant sidelobes.

\begin{figure}
\centering
\includegraphics[bb= 120 70 530 600,width=1.15\columnwidth]{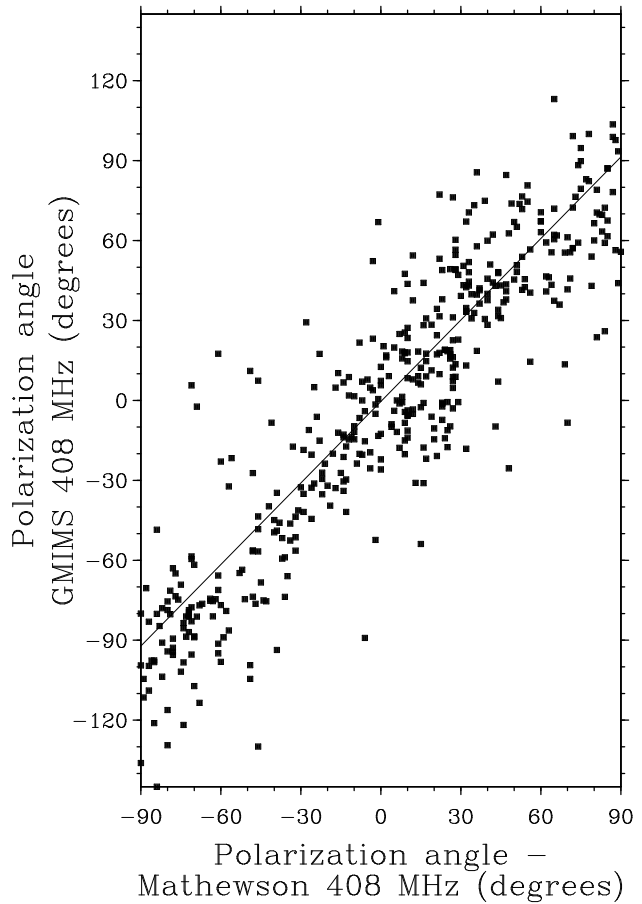}
 \caption{Comparison of polarization angle at 408\,MHz between the present work
 and the survey of \citet{math65}. The values for 437 points over the entire
 Southern sky common to the two surveys are shown. $60^{\circ}$ has been
 subtracted from the raw GMIMS values, and for clarity the GMIMS values have
 been permitted to fall outside the range ${\pm}90^{\circ}$. The solid line
 shows a fit to all points. The slope of the line is 1.02.}
 \label{math}
\end{figure}

\begin{figure}
\centering
\includegraphics[bb= 90 40 530 660,width=1.15\columnwidth]{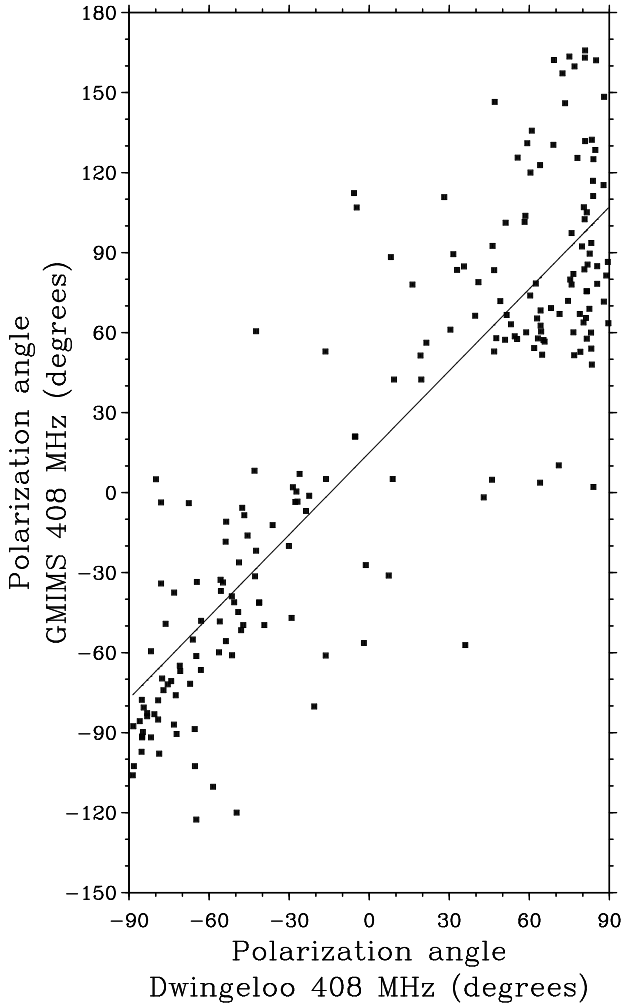}
 \caption{Comparison of polarization angle between the present work and the
 408\,MHz survey of \citet{brou76}. The comparison is confined to positions
 where polarized intensity exceeds 1\,K. The area of overlap is between
 declinations $0^{\circ}$ and $+20^{\circ}$, containing 193 common  points.
 $60^{\circ}$ has been subtracted from the raw GMIMS values, and the GMIMS
 values have been allowed to fall outside the range ${\pm}90^{\circ}$. The 
 solid line shows a fit to all points. The slope of the line is 1.02.}
 \label{dwing408}
\end{figure}

\begin{figure}
\centering
\includegraphics[bb= 90 40 530 660,width=1.15\columnwidth]{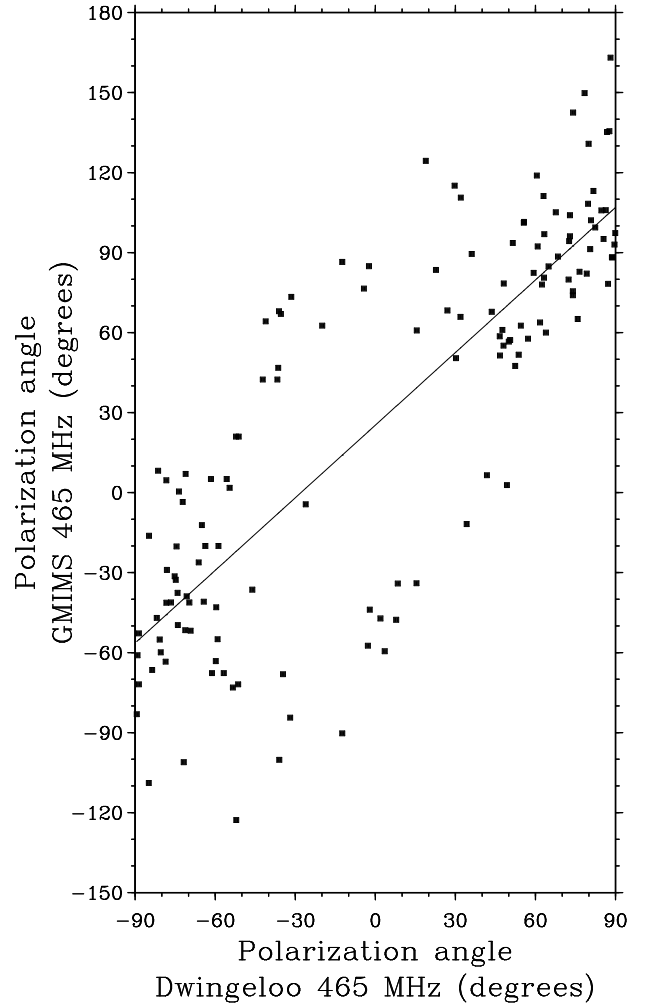}
 \caption{As for Figure~\ref{dwing408}, but for 465\,MHz. The plot contains 138
 points, and the slope of the fitted line is 0.91. See text for discussion.}
 \label{dwing465}
\end{figure}

In the face of this failure, we fell back on a post-observation comparison of
polarization angles with the data of \citet{math65}. We compared the two
datasets over most of the Southern sky, giving most weight to regions of high
polarized emission. From the comparison we deduced that a correction of
$-60^{\circ}$ should be applied to the GMIMS data, and this correction has been
applied to the released data. In Figure~\ref{math} we show the comparison after
the correction. We then made similar comparisons with the 408\,MHz and 465\,MHz
data from \citet{brou76}; these surveys extend only to declination $0^{\circ}$,
so the comparisons included fewer points. The results are shown in
Figures~\ref{dwing408} and \ref{dwing465}.

The data of \citet{math65} were obtained with a rotating dipole feed, for which
the zero is mechanically defined and should be absolute. However, this does not
ensure correct calibration of their final data because a substantial correction
for ionospheric Faraday rotation was applied and the angles were then adjusted
with measurements of reference regions taken from earlier data from the
Dwingeloo Telescope, data which subsequently went into the data published by
\citet{brou76}. In other words the comparisons in Figures~\ref{math} and
\ref{dwing408} are not entirely independent.  The fits illustrated by straight
lines plotted in Figures~\ref{math}, \ref{dwing408} and \ref{dwing465} may
indicate corrections slightly different from $-60^{\circ}$. However,  these fits
give equal weight to all data points; when the comparison is made with the
points where polarized intensity is highest, the offset of $-60^{\circ}$ emerges
very clearly. Although the comparisons were limited to 408 and 465\,MHz, there
does not seem to be a frequency dependent correction. While there is no further
check possible using independent data, our data alone do provide us with
assurance that the angles are close to correct. Inspection of the Faraday depth
cube after Rotation Measure Synthesis shows that, over most of the sky, the
strongest emission is at a Faraday depth of $0 \pm
0.5\,{\rm{rad}}\thinspace{\rm{m}}^{-2}$ and the all-sky average is very close to
$0\,{\rm{rad}}\thinspace{\rm{m}}^{-2}$, as would be expected.

\section{Results}
\label{results}

\begin{table*}
\caption{Available Data Products}
\label{products}
\begin{center}
\begin{tabular}{@{}lccccc}
\hline
Data Product & Beam  & Coverage & Freq. Resolution & FD Increment  & 
FD Coverage \\
 & (deg) & (MHz) & (MHz) & (${\rm{rad}}\thinspace{\rm{m}}^{-2}$)
 & (${\rm{rad}}\thinspace{\rm{m}}^{-2}$) \\
\hline
\hline
Total intensity & 1.35 & 286.25-487.75 & 0.5 & & \\
Stokes $Q$ and $U$ & 1.35 & 286.25-487.75 & 0.5 & & \\
Polarized intensity & 1.35 & 286.25-487.75 & 0.5 & & \\
Polarization angle & 1.35 & 286.25-487.75 & 0.5 & & \\
Faraday depth cube & & & & 0.5 & $-$100 to +100\\
\hline
\end{tabular}
\end{center}
\end{table*}

Detailed analysis of the survey data is beyond the scope of this paper, and in
this section we present only very general conclusions. The survey parameters are
summarized in Table~\ref{params}. The data from this survey are available from
the Canadian Astronomy Data Centre{\footnote{doi:10.11570/18.0007}.
Available data products are listed in Table~\ref{products}.  

The data products include total-intensity maps. They portray the sky as seen in
earlier surveys (in particular the survey of \citealp{hasl82}). We do not show
any total-intensity data here, as such a figure is unlikely to convey much new
information. The basketweaving process (see Section~\ref{basket}) inevitably
removes the sky minimum from the total-intensity data, so the zero level of
these images is not correct, but the intensity scale is absolutely calibrated at
all frequencies. What is new is the extensive frequency coverage, and we can
expect that the data do carry new information on spectral index, which might be
extracted with the T-T plot technique (see Figure~\ref{internaltt}). A detailed
study is beyond the scope of this paper. However, it is already evident from the
plots in Figure~\ref{internaltt} that the emission in this frequency range is
predominantly synchrotron emission. If significant thermal emission was present
one would expect to see a small branch on each T-T plot with a slope
corresponding to a lower spectral index. 

Compact sources in the survey
total-intensity data appear to have been broadened by about 20\% beyond the
beamwidths shown in Figure~\ref{beam}. This is probably the result of the
scanning strategy, where neighbouring scans, neeeded for accurate depiction of a
point source, may have been spaced by hours, days, or even months, and gain
drifts occurred over those time lapses. This point has been made about surveys
using scanning techniques like ours by \citet{reic88}.

\begin{figure}[!ht]
\centering
\includegraphics[bb = 1 10 330 500,width=0.9\columnwidth,angle=0,clip]
{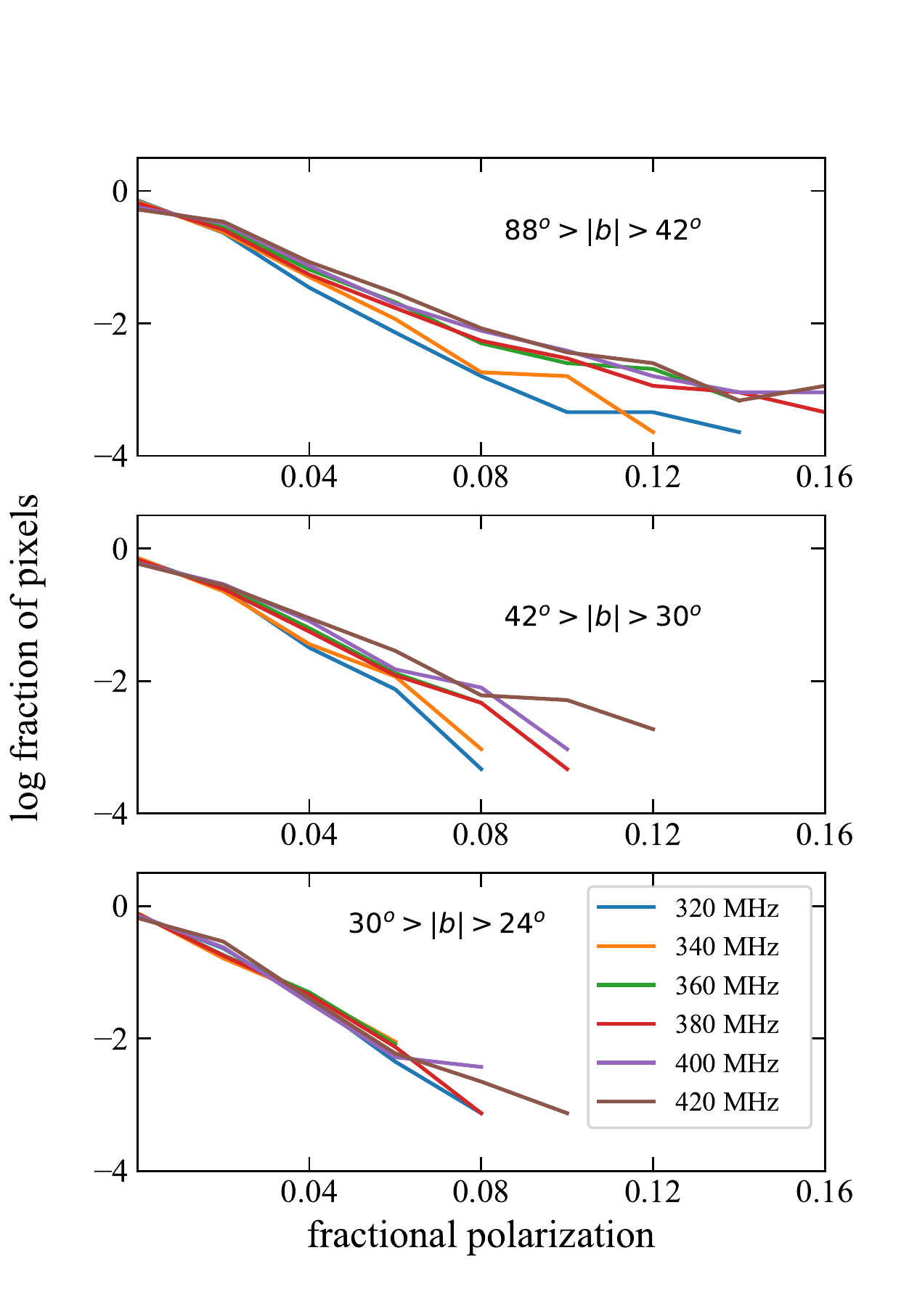}
 \caption{Histograms showing fractional polarization at frequencies across the
 band of the survey in three ranges of Galactic latitude. Calculation of 
 fractional polarization followed procedures described in the text.}
 \label{fract}
\end{figure}

Figure~\ref{fract} presents histograms of the fractional polarization over the
area of the survey, shown in three ranges of latitude.  To calculate these
values it was necessary to restore the zero level of the total-intensity maps.
At 400 MHz we added back the minimum sky brightness temperature from
\citet{hasl82}. At other frequencies we adjusted this by a factor appropriate
for ${\beta}={-2.5}$. Errors in these numbers will not seriously affect these
plots.  Latitude boundaries in Figure~\ref{fract} are chosen in steps of 0.5 in
${\rm{cosec}}|b|$. Assuming a plane-parallel magneto-ionic medium (MIM),
parallel to the Galactic plane, ${\rm{cosec}}|b|$ is the ratio of the path
length through the MIM to its scale height.

Three effects are evident in the data. First, fractional polarization is
generally very low: depolarization is strong along most sightlines. Second,
fractional polarization is slightly higher at higher frequencies. This can be
understood in terms of the polarization horizon, the maximum distance from which
polarized emission can be received at a particular frequency and beamwidth
\citep{uyan03}. Both depth depolarization and beam depolarization determine the
polarization horizon, and both are reduced at higher frequencies. Third,
independent of frequency, the highest latitude range displays slightly higher
fractional polarization, but only in a small number of directions. In a handful
of directions we may be `seeing' beyond the half-height of the MIM layer.

Artefacts appear in the $Q$ and $U$ images around a few strong sources ({\it{e.g.}} Virgo A, Hydra A, Fornax A, Pictor A). They are the product of cross-polarization in the feed and show the characteristic `four-leaf clover' pattern, alternating positive and negative lobes spaced $\frac{\pi}{2}$ around the source; the $Q$ and $U$ patterns differ by $45^{\circ}$. These lobes are at a level of a few percent. They arise from cross-polarization in the feed, converting $I$ into $Q$ and $U$. The same effect produces apparently polarized emission along the Galactic plane where total-intensity emission is very strong
(see Figures~\ref{moments} and \ref{fd}). This too is spurious.

\begin{figure*}
\centering
\includegraphics[bb= 60 60 600 530, width=1.4\columnwidth,angle=0,clip=]{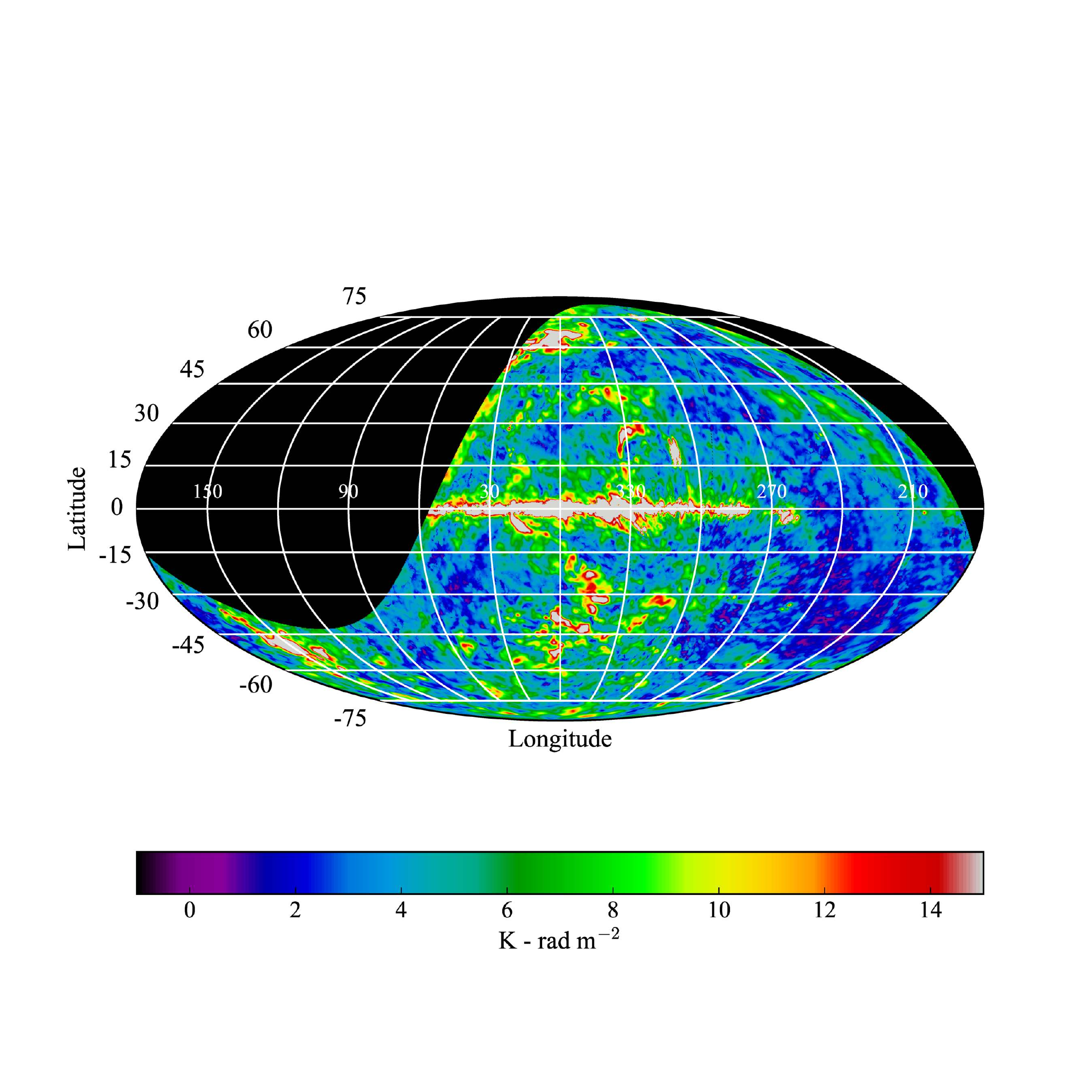}
\includegraphics[bb= 60 60 600 530, width=1.4\columnwidth,angle=0,clip=]{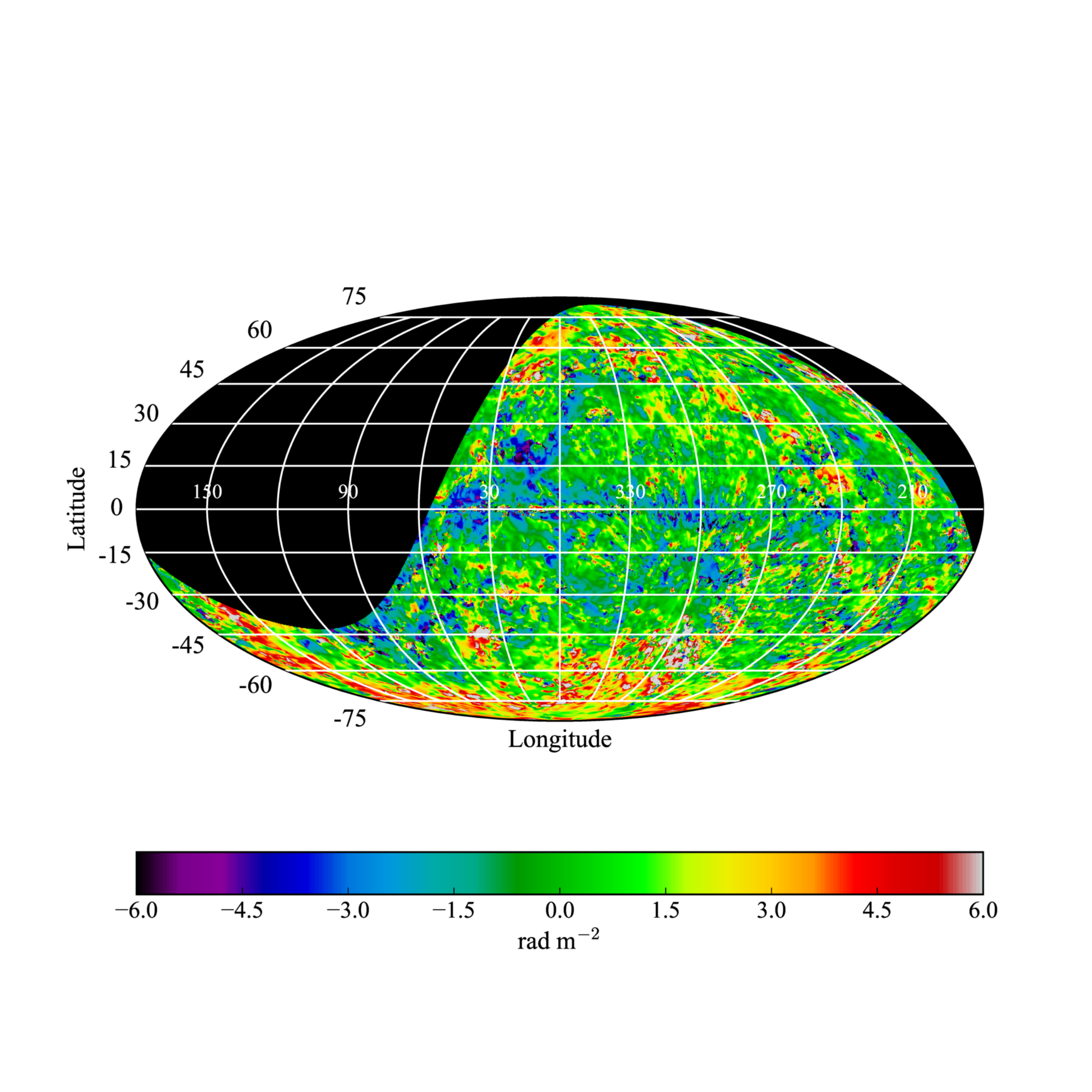}
 \caption{Zero moment map (top) and first moment map (bottom) calculated from the Faraday depth cube. Both are in Galactic coordinates, shown in Mollweide projection.}
 \label{moments}
\end{figure*}

\begin{figure*}
\centering
\includegraphics[bb= 60 60 600 530, width=1.4\columnwidth,clip=]{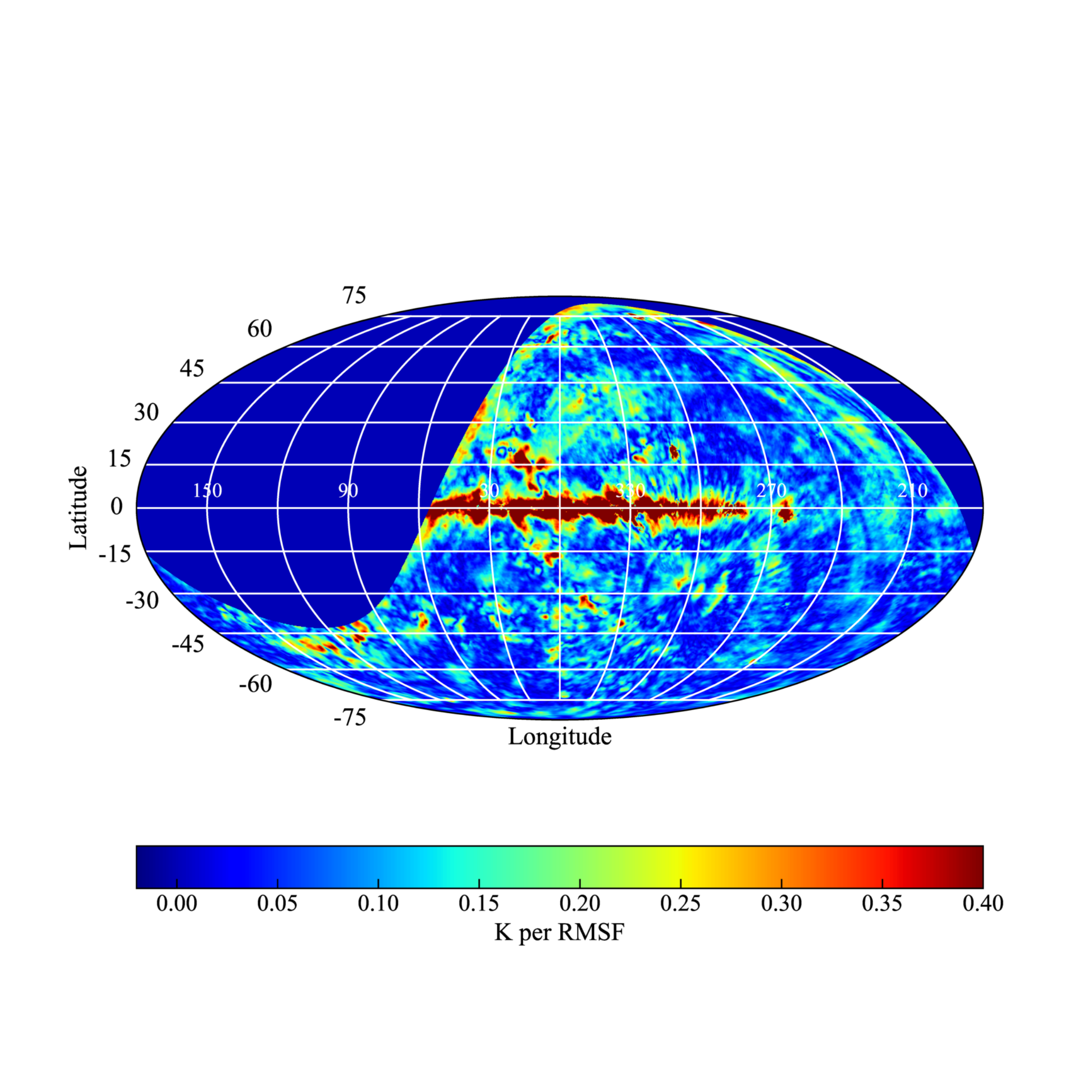}
\includegraphics[bb= 60 60 600 530, width=1.4\columnwidth,clip=]{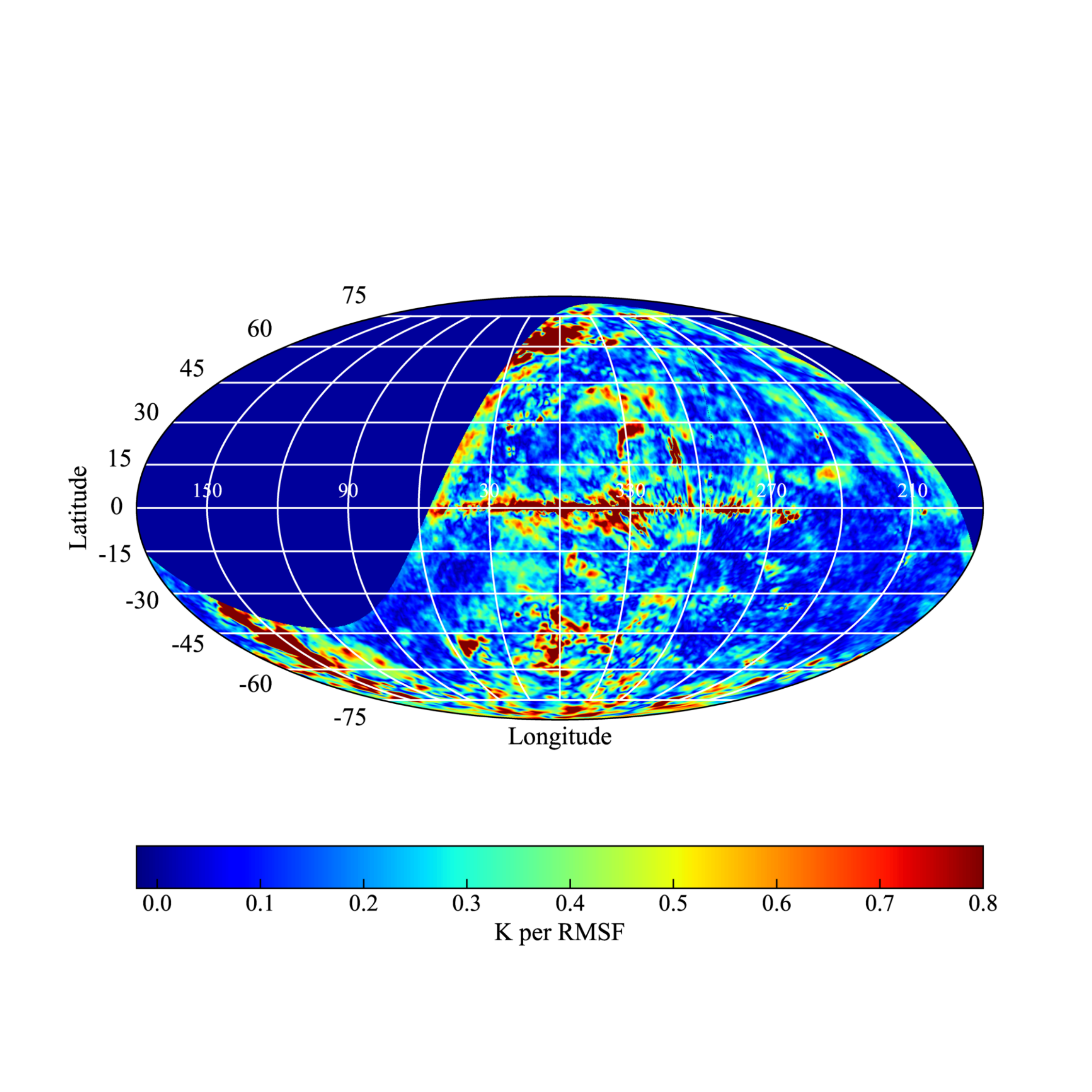}
 \caption{Faraday depth images at values of $-$8.5 (top) and
$+5\,\rm{rad}\thinspace{\rm{m}}^{-2}$ (bottom). Note different intensity scales for the two plots. Both are in Galactic coordinates, shown in Mollweide projection.}
 \label{fd}
\end{figure*}

\begin{figure*}
\centering
\includegraphics[width=2.0\columnwidth,angle=0]{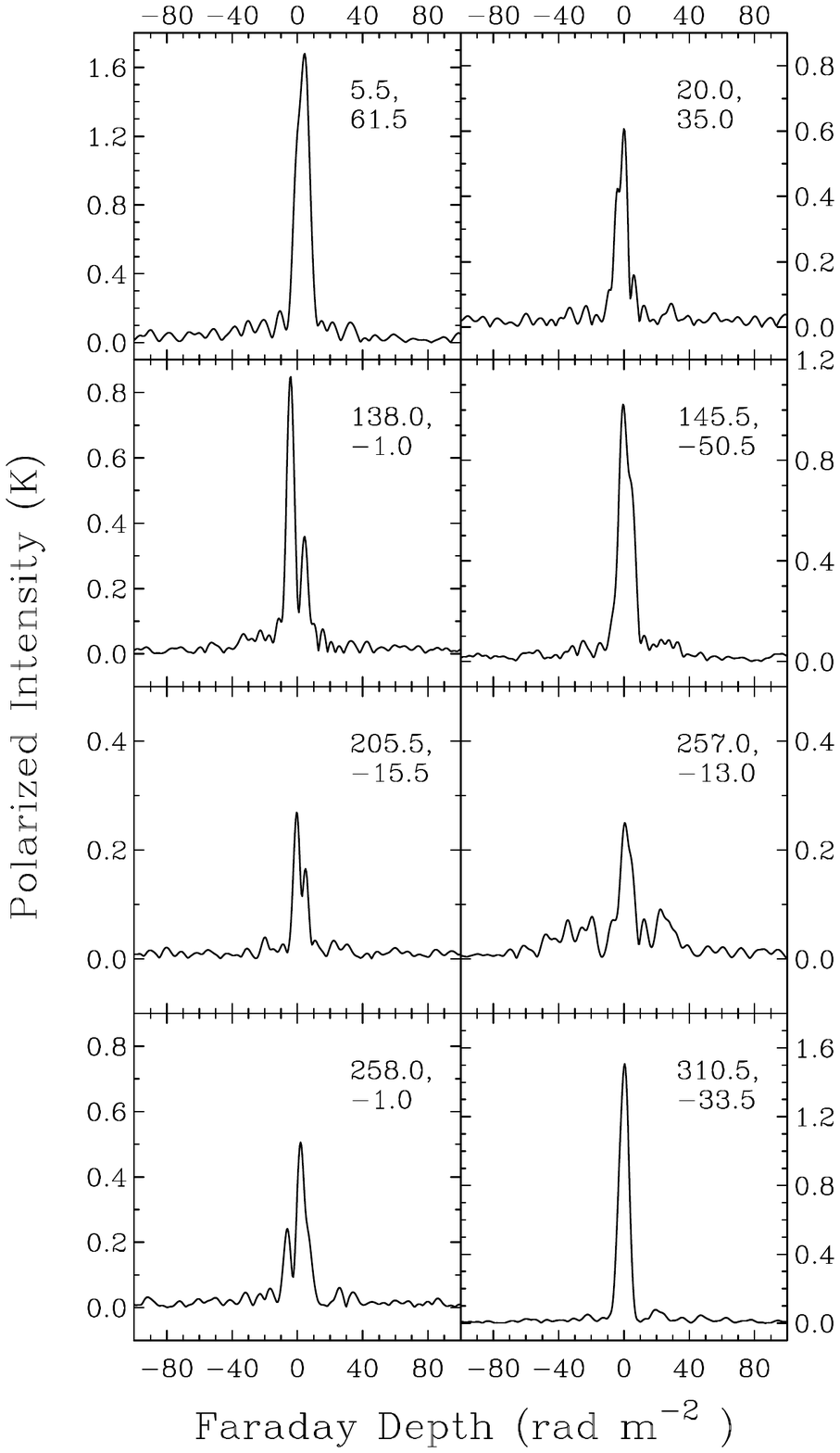}
 \caption{Faraday depth spectra. Sky position in Galactic co-ordinates is shown
 for each spectrum. Different vertical scales are used for some positions.}
 \label{rm_spectra}
\end{figure*}

The data product from this survey that is without precedent is the Faraday-depth
cube.  \citet{dick19} calculate the moments of Faraday depth from this cube,
from order zero to order two. Here we present images of the zeroth and first
moments. The zeroth moment (top image in Figure~\ref{moments}) shows the total
polarized brightness integrated over the full range of Faraday depth. This is a
representation of the locations of bright polarized emission without presenting
maps at specific frequencies. This image bears a strong resemblance to the map
of polarized intensity at 408\,MHz shown by \citet{math65}. The lower image of
Figure~\ref{moments} is the first moment map, which, for every pixel, shows
the weighted average value of Faraday depth in that direction; in directions
where the Faraday depth spectrum is simple this is close to a map of peak
Faraday depth. Figure~\ref{fd} shows images at Faraday depths of $-8.5$ and
$+5\,{\rm{rad}}\thinspace{\rm{m}}^{-2}$.

Figure~\ref{rm_spectra} presents a few characteristic Faraday depth spectra from
the survey. We do not attempt interpretation here, but simply point out the
existence of multiple Faraday-depth components at some positions, and the fact
that significant emission is found at non-zero values of Faraday depth. In some
of these directions the true Faraday depth spectrum may be a single wide peak:
it is possible that our survey sees two peaks because we are not sensitive to
Faraday depth structures wider than $8.6\,{\rm{rad}}\thinspace{\rm{m}}^{-2}$
(see Table~\ref{params}). \citet{schn09} similarly report complex Faraday
spectra seen in high-resolution (${2.8'}\times{4.7'}$) data covering 324 to
387\,MHz. We note the very complicated spectrum at
${\ell,b}={257.0^{\circ},13.0^{\circ}}$: this position is on the southern part
of the Gum Nebula \citep{purc15}.

Given the significant problems that we encountered with RFI, the survey data
products are not noise limited. Scanning artefacts remain in the images, and
they are probably attributable to remnants of low-level RFI. Fluctuations on
images are at levels of 1\,K rms in total-intensity maps and 120\,mK rms in $Q$
and $U$ maps at a beamwidth of $1.35^{\circ}$. Fluctuations on images in the
Faraday depth cube are $\sim$20\,mK in single channels of width
$0.5\,{\rm{rad}}\thinspace{\rm{m}}^{-2}$.

\section{Concluding Discussion}
\label{disc}

We have described a large survey of polarized emission, covering 67\% of the
visible sky. We have demonstrated that the rapid azimuth scanning technique
developed by Carretti for S-PASS is a viable technique for making polarization
observations of the linearly polarized sky at low frequencies. The intensity
scale is absolutely calibrated at all frequencies, and is accurate to 7\%.
Processing has concentrated on the extended emission; users of the data should
be aware that some compact sources may not be accurately portrayed. Our choice of  parameters for Rotation Measure synthesis (Section~\ref{rot_meas_synth}) may not suit every application. We encourage users of the data to employ their own routines, or to apply improved routines which may be developed in the future.

This is the first of the GMIMS surveys to reach publication, providing all-sky
coverage of the diffuse Galactic emission with the application of Rotation
Measure Synthesis. The resolution of the survey in Faraday depth is
$5.9\,\rm{rad}\thinspace{\rm{m}}^{-2}$, and the survey is capable of capturing
Faraday depth features with a width of $8.6\,\rm{rad}\thinspace{\rm{m}}^{-2}$.
Our results demonstrate that there is emission at non-zero Faraday depths. Many
directions display Faraday depth spectra with multiple peaks, commonly two, but
sometimes more. There is polarized emission beyond the reach of single-frequency
surveys. Had we been able to use data up to 900\,MHz we would have been able to
``see'' structures as wide as ${\sim}30\,\rm{rad}\thinspace{\rm{m}}^{-2}$: good
resolution in Faraday depth requires observing to long wavelengths, and ability
to image broad Faraday depth features requires broad coverage in wavelength.

Wideband observations like this survey necessarily stray outside the traditional
radio astronomy frequency allocations, which are in any event quite narrow. From
our experience we have learned that successful observing is possible, even under
conditions of quite intense RFI. The key to our success was heavy spatial
oversampling of the sky. Because the survey was planned for full sampling at
900\,MHz we observed every point of the 300\,MHz sky on the order of ten times.
This proved to be sufficient to achieve a very high fractional coverage of the
survey area. This has been achieved, of course, at the expense of valuable
telescope time and observer time, but there seems to be no simple alternative.
We note that this kind of oversampling can be successful as long as the RFI is
intermittent, even if band occupancy is high. Nevertheless, we must acknowledge
that our images are not noise limited, they are artefact limited and the
artefacts are probably RFI at levels below our excision process. We cut off the
RFI excision at a level where we judged that the data contained valid and useful
astrophysical information.

In our case astrophysical requirements would have dictated a channel width no
narrower than a few MHz, but using narrower bandwidths meant that data affected
by RFI could be deleted without losing a significant amount of good data. It is
best if the channel width matches the width of the RFI signals, and  our use of
0.5\,MHz channels came close to this. Current digital data processing systems
are capable of delivering narrow channel widths, and they should be used.

Our survey is absolutely calibrated in amplitude. This was necessary because we
observed well away from traditional radio astronomy bands and there was little
data that we could use to transfer calibration. The fact that our data is
published in units of Kelvins of brightness temperature is important to allow us
to link this survey to other GMIMS data products. The intensity scale compares
well at 408\,MHz with the scale of the survey of \citet{hasl82}. Beyond that
comparison, only internal checks on the amplitude scale are possible. They
depend on assumptions about the spectral index of synchrotron emission, and,
because the frequency range is narrow, they have limited precision. The weakest
point of our calibration is the calibration of polarization angle, and we have
in the end been compelled to use the excellent data of \citet{brou76}.
Nevertheless, the most important data product from our survey is the Faraday
depth cube: for many purposes the actual polarization angles are less important
than the complex change of angle as a function of frequency.

The data presented in this paper are available at the Canadian Astronomy Data Centre at doi:10.11570/18.0007.

\acknowledgments

Rob Messing, Richard Hellyer and Ev Sheehan of DRAO put skill and hard work into
building the receiver with us. The Parkes Radio Telescope is part of the
Australia Telescope National Facility which is funded by the Commonwealth of
Australia for operation as a national facility managed by CSIRO. We could not
have started this survey without the help of the staff at the Parkes
Observatory, and their endless enthusiasm helped us bring it to fruition: John
Reynolds, Brett Armstrong, Ken Reeves and Mal Smith helped us extensively with
receiver and RFI issues. Mike Kesteven contributed to the development of the
telescope drive software that made fast scanning possible. Jana Koehler, Stasi
Baran, Niloofar Gheissari, Xiaohui Sun, LiGang Hou, and Dominic Schnitzeler
helped with observations. In processing the survey data we used the facilities
of the Centre for High Performance Computing in Capetown, South Africa.
Tony Willis provided vauable help with calculations of ionospheric Rotation
Measure. The Max-Planck-Institut f\"ur Radioastronomie provided some essential
components for the receiver used in this work. MW was supported in part by funds
from the Natural Sciences and Engineering Research Council. AF thanks the STFC
and the Leverhulme Trust for financial support. The Dunlap Institute is funded
through an endowment established by the David Dunlap family and the University
of Toronto. Sean Dougherty and Roland Kothes were generous in their support of
this project from outset to completion. And finally, this paper has
benefited greatly from a thorough review by a knowledgable referee.

\bibliographystyle{aasjournal}
\bibliography{parkes_arxiv} 
\end{document}